\documentclass[preprints,review,accept,pdftex,moreauthors]{mdpi} 
\firstpage{1} 
\pubvolume{1}
\issuenum{1}
\articlenumber{1}
\pubyear{2023}
\copyrightyear{2023}
\datereceived{Feb.~9, 2023} 
\dateaccepted{Mar.~20, 2023} 
\datepublished{Apr.~14, 2023} 
\hreflink{https://doi.org/\linebreak10.3390/galaxies11020056} 
\doinum{10.3390/galaxies11020056}

\Title{Convective boundary mixing in main-sequence stars: theory and empirical constraints}

\TitleCitation{Convective boundary mixing in main-sequence stars: theory and empirical constraints}

\Author{Evan H. Anders $^{1,\dagger,*}$\orcidA{} and May G. Pedersen $^{2,3,\dagger}$\orcidB{}}

\AuthorNames{Evan H. Anders and May G. Pedersen}

\AuthorCitation{Anders, E. H.; Pedersen, M. G.}

\address{%
$^{1}$ \quad 
Center for Interdisciplinary Exploration and Research in Astrophysics (CIERA), Northwestern University, 1800 Sherman Ave, Evanston, IL 60201,USA; evan.anders@northwestern.edu
\\
$^{2}$ \quad Sydney Institute for Astronomy, School of Physics, University of Sydney, Sydney, NSW 2006, Australia; may.pedersen@sydney.edu.au\\
$^{3}$ \quad Kavli Institute for Theoretical Physics, Kohn Hall, University of California, 
Santa Barbara, CA 93106, USA
}

\corres{Correspondence: evan.anders@northwestern.edu}

\firstnote{These authors contributed equally to this work.} 

\abstract{
    The convective envelopes of solar-type stars and the convective cores of intermediate- and high-mass stars share boundaries with stable radiative zones.
    Through a host of processes we collectively refer to as ``convective boundary mixing'' (CBM), convection can drive efficient mixing in these nominally stable regions.
    In this review, we discuss the current state of CBM research in the context of main-sequence stars through three lenses.
    (1) We examine the most frequently implemented 1D prescriptions of CBM---exponential overshoot, step overshoot, and convective penetration---and we include a discussion of implementation degeneracies and how to convert between various prescriptions.
    (2) Next, we examine the literature of CBM from a fluid dynamical perspective, with a focus on three distinct processes: convective overshoot, entrainment, and convective penetration.
    (3) Finally, we discuss observational inferences regarding how much mixing should occur in the cores of intermediate- and high-mass stars, and the implied constraints that these observations place on 1D CBM implementations.
    We conclude with a discussion of pathways forward for future studies to place better constraints on this difficult challenge in stellar evolution modeling.
}

\keyword{UAT keywords: Stellar Evolution (1599), Stellar Evolutionary Models (2046), Stellar Convection Zones (301), Stellar Cores (1592), Hydrodynamical Simulations (767), Star Clusters (1567), Apsidal Motion (62), Asteroseismology (73), Stellar Oscillations (1617), Binary Stars (154)} 

\graphicspath{{./figures/}}
\newcommand{\yL}{\ensuremath{\mathcal{Y}_{\rm{L}}}}
\newcommand{\yS}{\ensuremath{\mathcal{Y}_{\rm{S}}}}
\newcommand{\gradrad}{\ensuremath{\nabla_{\rm{rad}}}}
\newcommand{\gradad}{\ensuremath{\nabla_{\rm{ad}}}}

\newcommand{\gradmu}{\ensuremath{\nabla_{\mu}}}

\newcommand{\grad}{\ensuremath{\nabla}}
\renewcommand{\vec}{\ensuremath{\boldsymbol}}
\newcommand{\angles}[1]{\ensuremath{\left\langle #1 \right\rangle}}
\newcommand{\brunt}{Brunt-V\"ais\"al\"a}
\newcommand{\Ri}{\ensuremath{\mathrm{Ri}_B}}
\newcommand{\mS}{\ensuremath{\mathcal{S}}}

\usepackage{ulem}
\usepackage{xcolor}
\newcommand{\edit}[2]{\textcolor{orange}{\textbf{#2}}}

\begin{document}


\section{Introduction}

Convection occurs in all main-sequence stars, and there is broad agreement that widely-used prescriptions like the mixing length theory \citep[MLT, ref.~][discussed in chapter 1 in this series]{bohm-vitense1958} adequately describe many properties of bulk convection in stellar interiors. 
There is, however, a great deal of disagreement and uncertainty regarding how to model the boundaries of convection zones, where the stellar stratification changes from being convectively unstable to stable.
Convective boundaries exist at radial coordinates where \emph{the buoyant force} changes sign (from acceleration to deceleration \citep{anders_etal_2022b}), but most models and MLT unphysically assume that \emph{the convective velocity} vanishes at these locations.
The true boundary of a convection zone---the location where the convective velocity is zero---lies beyond the traditional ``convective boundary,'' and some parameterization of ``convective boundary mixing'' (CBM) is generally employed alongside MLT to allow convective motions to extend outside of the MLT convection zone.

While low-mass stars are fully convective, stars like the Sun with masses $0.35 M_{\odot} \lesssim M_* \lesssim 1.2 M_{\odot}$ develop stable interiors and turbulent convective envelopes \citep{charbrier_baraffe_1997,chabrier_baraffe_2000,jermyn_etal_2022_atlas}.
Convective motions can ``undershoot'' from the unstable envelope into the stable interior and cause mixing, which can alter surface Lithium abundances \citep{pinsonneault_1997,carlos_etal_2019,dumont_etal_2021} and the sound speed below the convection zone \citep{christensen-dalsgaard_etal_2011,bergemann_serenelli_2014,basu_2016}.
In ``massive stars'' with masses $M_* \gtrsim 1.1 M_{\odot}$, efficient nuclear burning from the CNO cycle destabilizes the core to convection, while the envelope becomes convectively stable  \citep{hansen_etal_2004,jermyn_etal_2022_atlas}; some of these stars also have opacity-driven convection zones near the surface \citep{cantiello_braithwaite_2019,jermyn_etal_2022_window}.
CBM from the core convection zone injects fresh hydrogen fuel into the high-temperature central burning region of the star, thereby extending the stellar lifetime and increasing the size of the helium core at the end of the main sequence.
Unfortunately, observations cannot be uniformly explained with one standard CBM prescription \citep{johnston_2021}, leading to uncertainty in how to include CBM in stellar evolution calculations.
These uncertainties are not subtle: evolving a 15 $M_{\odot}$ model using differing mixing prescriptions can significantly alter the main sequence lifetime by $\sim 25\%$ and the helium core mass by $\sim 40\%$, with consequences that ripple beyond the main sequence, such as in determining what type of remnant the star eventually leaves behind \citep{kaiser_etal_2020,schootemeijer_etal_2019}.
Fortunately, there seems to be a tight relationship between a star's core mass, its envelope mass, and its core composition if the star is to remain in equilibrium \citep{farrell_etal_2020}, which may limit the range of feasible CBM prescriptions to characterize.

Observations of massive stars cannot be explained without CBM which increases the convective core size compared to ``standard'' stellar models.
For example, radial profiles of the \brunt~frequency measured via asteroseismology demonstrate substantial mixing outside the standard core boundary \citep{Pedersen2021NatAs...5..715P}.
The population of observed eclipsing binaries \citep{Claret2018ApJ...859..100C} and the width of the main sequence in the temperature-luminosity plane observed in stellar clusters \citep{castro_etal_2014,higgins_vink_2019,martinet_etal_2021,higgins_vink_2023} can be partially explained by introducing a mass-dependent CBM into stellar models.
Simulations of 3D turbulent convection whose initial conditions are based on 1D stellar evolution models consistently report significant entrainment at convective boundaries and expansion of the convection zone (e.g.,~\citep{meakin_arnett_2007,gilet_etal_2013,cristini_etal_2017,jones_etal_2017,andrassy_etal_2020,higl_etal_2021,rizzuti_etal_2022}), so the picture from numerical simulations aligns with that inferred from observations.

In this review, we discuss CBM in stars.
Our goals in writing this review are:
\begin{enumerate}
    \item to provide context for investigators who need to employ CBM in their own studies.
    \item to summarize past works and provide launching points for future studies aimed at improving CBM prescriptions.
    \item to facilitate communication between observers, 1D modelers, and 3D numericists.
\end{enumerate}
In section~\ref{sec:parameterizations}, we describe 1D parameterizations of CBM.
In section~\ref{sec:simulations}, we describe the results of numerical simulations exhibiting CBM.
In section~\ref{sec:observations}, we describe observations and empirical calibrations of CBM.
We conclude with a discussion in section~\ref{sec:discussion}.


\section{Theoretical (1D) parametrizations}
\label{sec:parameterizations}

\subsection{How does CBM modify stellar evolution?}
\label{subsec:what_is_cbm}

In stellar evolution software instruments, the mixing caused by convection, CBM, and other mixing processes are generally parameterized as a turbulent diffusivity \citep{brandenburg_etal_2009}. 
That is, for some chemical composition (e.g., hydrogen, $X$), time evolution is assumed to follow,

\begin{equation}
    \partial_t X = \grad\cdot [(D_{\rm conv} + D_{\rm CBM} + D_{\rm other})\grad X],
\end{equation}

when changes to the composition from nuclear reactions are ignored. 
Here each $D$ is a diffusion coefficient.
In this formalism, it is impossible to distinguish between CBM and other mixing processes which could occur in the vicinity of a convective boundary (e.g., shear, rotation, etc).
This formalism generally allows us to probe the \emph{shape} and \emph{magnitude} of the radial profile of mixing, but not the fundamental process at work.
Regardless, within this review we will assume that excess mixing which connects to and extends beyond the convective boundary are convection-induced CBM processes.
We additionally limit the scope of this review to purely hydrodynamical CBM processes; complicating effects of e.g., magnetism or radiative transfer are not considered. 

\begin{figure}[t]
\includegraphics[width=\linewidth]{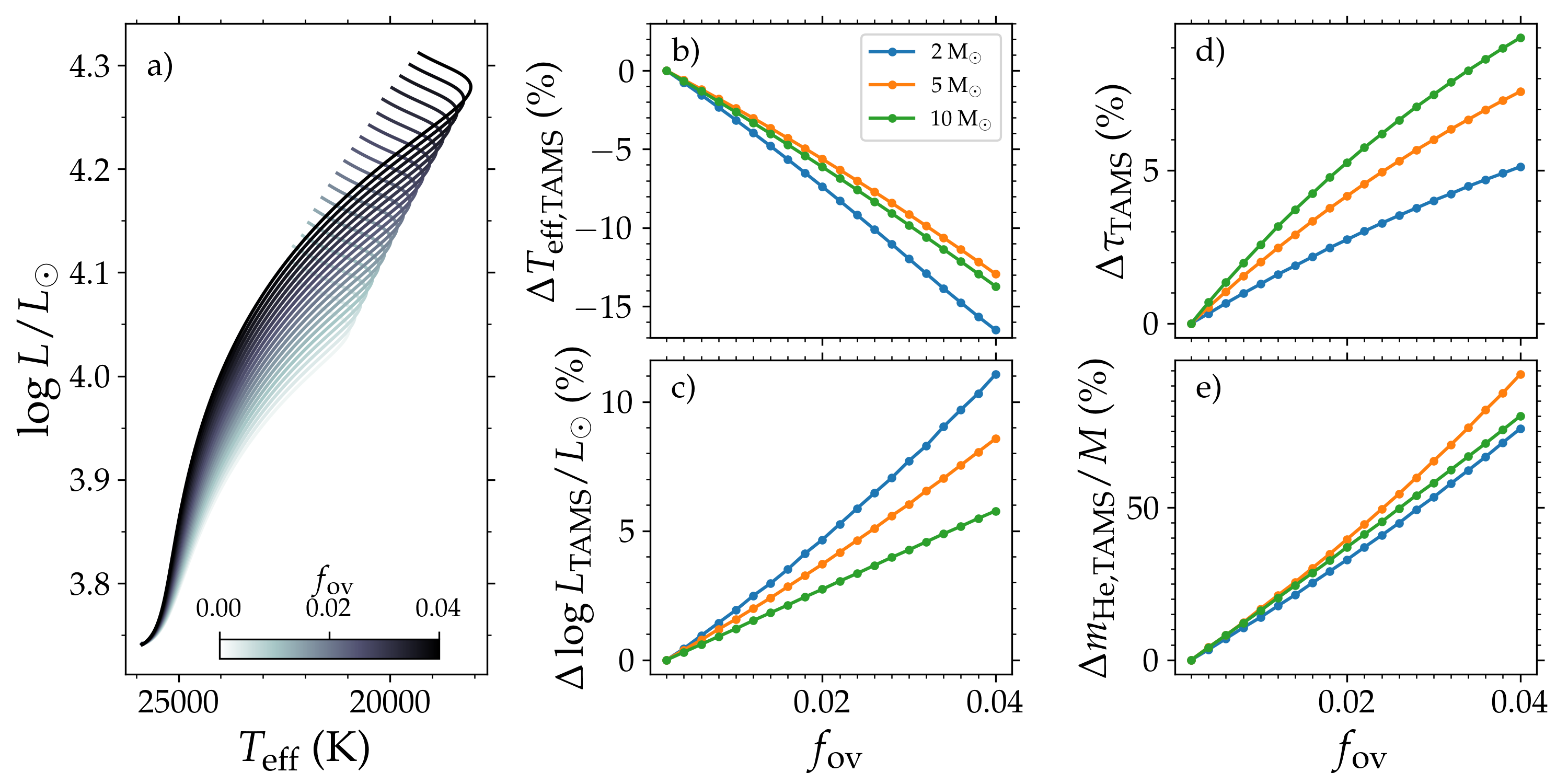}
\caption{Illustration of the effect of increasing the size of the CBM region assuming exponential diffusive overshoot. 
(\textbf{a}) HR diagram showing the evolutionary tracks for a 10\,M$_\odot$ star with different extents of the CBM region. 
The $f_\text{ov}$ parameter sets the extent of the CBM region and how rapidly the mixing decreases with distance from the convective core boundary, see Section~\ref{subsec:Overshoot}. 
$f_\text{ov} = 0.002$ is a low value, whereas $f_\text{ov} = 0.04$ is considered a high value for this parameter. 
(\textbf{b}) Percentage change in the effective temperature at the TAMS compared to the $f_\text{ov} = 0.002$ case for three different initial masses. 
(\textbf{c}) Same as (\textbf{b}) but for the luminosity at the TAMS. (\textbf{d}) Same as (\textbf{b}) but for the age at the TAMS. (\textbf{e}) Same as (\textbf{b}) but for the helium core mass obtained at the TAMS. Figure made by the authors using \texttt{MESA} models. \texttt{MESA} inlists and data used to generate the figure are available on Zenodo \citep{supp}. 
\label{fig:CBM_effects}}
\end{figure}

In Figure~\ref{fig:CBM_effects}, we briefly demonstrate how CBM affects the evolution of stars with convective cores.
Panel a shows that increasing the radial extent of mixing beyond the convective boundary (going from light to darker lines) decreases a star's effective temperature (panel b) and increases its luminosity (panel c) at the TAMS (terminal-age main sequence).
This increased mixing also increases the length of the main sequence (panel d) by providing more fuel and significantly increases the helium core mass at the end of the main sequence (panel e); these latter changes introduce uncertainty into the star's post-main-sequence evolution and into the eventual remnant that the star leaves behind.
While not pictured here, vectors in the mass--luminosity plane can help disentangle the effects of mass loss and internal mixing on the evolutionary tracks of very massive stars \cite{higgins_vink_2019,higgins_vink_2023}. 
The slope of the vector is set by the mass-loss rate, while the length of the vector is set by the age and internal mixing, see Figure~1 in \citet{higgins_vink_2023}. 
If the star's rotational velocity is known, the CBM parameter can be directly derived from the length of the vector. 

\subsection{Convective boundaries}
\label{subsec:convboundaries}

In 1D stellar evolution software instruments, convection zone boundaries coincide with a sign change in a determinant $\mathcal{Y}$ (\citep{Paxton2018ApJS..234...34P}, Section 2).
We define regions with $\mathcal{Y} < 0$ to be \emph{stable} to convection.
The simplest convective stability criterion is the Schwarzschild criterion,
\begin{equation}
    \yS \equiv \gradrad - \gradad < 0.
    \label{eqn:yS}
\end{equation}
Here, the logarithmic temperature gradient is $\grad \equiv d \ln P / d \ln T$ (pressure $P$ and temperature $T$).
When $\grad$ is evaluated at constant entropy and mean molecular weight, its value is the adiabatic gradient $\gradad$.
The gradient required to radiatively transport the full stellar luminosity is $\gradrad$.

In the presence of gradients in the mean molecular weight $\mu$, a better convective stability criterion is the Ledoux criterion,
\begin{equation}
    \yL \equiv \yS +  \frac{\chi_\mu}{\chi_T}\gradmu < 0.
    \label{eqn:yL}
\end{equation}
The Ledoux criterion includes the composition gradient $\gradmu = d\ln\mu/d\ln P$, where $\chi_T = (d\ln P / d\ln T)_{\rho,\mu}$, $\chi_\mu = (d\ln P / d\ln\mu)_{\rho,T}$, and $\rho$ is the density.
The composition term is generally stabilizing when the radial gradient of $\mu$ is negative.

In this review, we are interested in cases where an unstable convection zone (CZ) with $\yL \geq 0$ borders a stable radiative zone (RZ) with $\yL \leq \yS < 0$.
We therefore do not consider e.g., semiconvection or thermohaline mixing (see chapters 2 \& 3 of \citet{garaud_2018} as well as section 3 \& figure 3 of \citet{salaris_cassisi_2017}).
We note that evolutionary timescales are much longer than the convective overturn timescale on the main sequence \citep{georgy_etal_2021}.
In this regime, both the Ledoux criterion and the Schwarzschild criterion should retrieve the same location for the convective boundary, as argued by \citet{gabriel_etal_2014} and shown using hydrodynamical simulations by \citet{anders_etal_2022b}.
Therefore, we will refer to the convective boundary and the Schwarzschild boundary interchangeably.

The Schwarzschild boundary ($\yS = 0$) generally corresponds to an interface where the entropy gradient goes from being marginally stable (or unstable, $\grad s \leq 0$) to being stable ($\grad s > 0$).
``Convective boundaries'' defined by $\mathcal{Y}$ therefore specify where the radial buoyancy force changes from destabilizing (in the convection zone) to stabilizing (in the radiative zone).
The location where convective velocity goes to zero therefore always lies ``outside'' of the convective boundaries defined by $\yS$.
The CBM prescriptions that we discuss below therefore attempt in spirit to estimate the size of the region in which convective velocities decelerate beyond the convective boundary.


\subsection{Internal mixing profiles}
\label{subsec:mixing_profs}

The time evolution of the mass fraction $X_i$ of chemical element $i$ depends on nuclear reactions $\mathcal{R}_i$ and mixing processes $\mathcal{M}_i$. 
1D stellar evolution software instruments typically treat element mixing as a diffusive process, so the time evolution equation 
\begin{align}
    \frac{\partial X_i}{\partial t} &= \mathcal{R}_i  + \mathcal{M}_i\nonumber, \\
    &= \mathcal{R}_i + \frac{1}{\rho r^2} \frac{\partial}{\partial r} \left[ \rho r^2 D_{\rm mix}  \frac{\partial X_i }{\partial r}\right] + \mathcal{M}_i^{\rm micro},
\end{align}
where $\rho$ is the density and $r$ is the radius coordinate. 
We group extra microscopic atomic diffusion effects like radiative levitation or gravitational settling into $\mathcal{M}_i^{\rm micro}$. 
The mass fraction $X_i$ diffuses with a diffusivity of $D_{\rm mix}$ in units of cm$^2$\,s$^{-1}$. 

\begin{figure}[t]
\includegraphics[width=\linewidth]{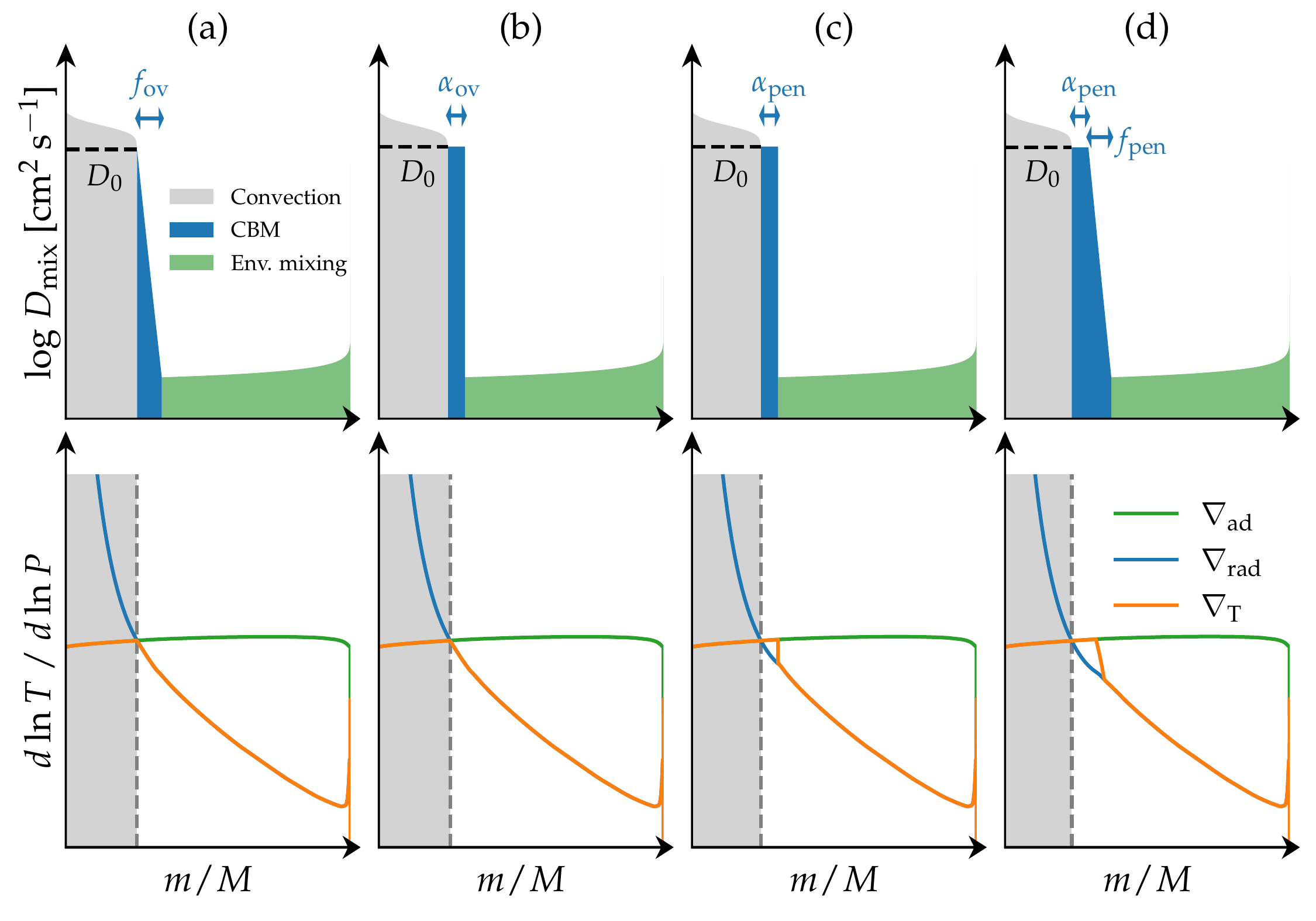}
\caption{Illustration of four different convective boundary mixing prescriptions in 1D. Internal mixing profiles are shown in the top panels, while the corresponding temperature gradients are given in the bottom panels. (\textbf{a}) Exponential diffusive overshoot, (\textbf{b}) step overshoot, (\textbf{c}) convective penetration, and (\textbf{d}) extended convective penetration. Convective regions are indicated in grey. $D_{\rm CBM} (r)$ and $D_{\rm env} (r)$ are shown respectively in blue and green in the top panels. In the bottom panels, the plotted curves show the adiabatic temperature gradient (green), the radiative temperature gradient (blue), and the adopted temperature gradient (orange). Figure made by the authors using \texttt{MESA} models. The data used to generate the figure are available on Zenodo \citep{supp}.
\label{fig:CBM_1D}}
\end{figure}

The sum of a variety of different physical process such as convection, rotation, magnetic fields, and waves all contribute to $D_{\rm mix} (r)$ in different regions and at different magnitudes. 
We decompose the turbulent diffusivity based on whether or not convection is present,
\begin{equation}
    D_{\rm mix} (r) = D_{\rm conv} (r) + D_{\rm CBM} (r) + D_{\rm env} (r),
\end{equation}
where $D_{\rm conv} (r)$ is the contribution from convective regions, $D_{\rm CBM} (r)$ characterizes convective boundary mixing regions, and $D_{\rm env} (r)$ is the mixing profile in the radiative envelope. 
Parameterizing mixing into diffusion profiles in this way discards information about the specific processes that cause mixing, which makes it difficult to disentangle the individual mixing contributions of CBM, rotational mixing, and other processes that occur at the same radial coordinate.
Examples of four $D_{\rm mix} (r)$ profiles are illustrated in the top panels of Figure~\ref{fig:CBM_1D}. 
The associated stratification produced by these mixing coefficients is shown in the bottom panels of Figure~\ref{fig:CBM_1D}.

The remainder of this section will focus on the different $D_{\rm CBM} (r)$ parameterizations that are commonly used in 1D stellar evolution codes. 
We use the term CBM to refer to \emph{any} boundary mixing process. 
We adopt a terminology where \emph{convective overshoot} only mixes chemical composition so that  $(\nabla_\text{T} = \nabla_\text{rad})$ in the CBM region, and \emph{convective penetration} mixes both chemical composition and entropy so that  $(\nabla_\text{T} = \nabla_\text{ad})$  in the CBM region. 

\edit{
    Aside from the prescriptions discussed below, other CBM prescriptions such as the diffusive Gumbel overshooting \citep{Pratt2017A&A...604A.125P,Augustson2019ApJ...874...83A} and diffusive double exponential overshoot \citep{Herwig2007ASPC..378...43H} profiles have been proposed.
    We are unaware of any cases where these prescriptions have been tested and compared against observations, so we exclude them from the following discussion.
    There have been other attempts to derive and calibrate convective boundary mixing prescriptions from first-principles, such as ``turbulent convection models'' \citep[e.g.,][]{zhang_li_2012,zhang2016}, Canuto's stellar mixing model \citep[e.g.,][]{canuto2011}, and non-local ``turbulent kinetic energy models'' \citep[e.g,.][]{kupka_montgomery_2002,kupka_etal_2022}.
    We focus on the most frequently-used techniques in the literature, so a discussion of these models is beyond the scope of this work.
    }{}

\subsection{Overshoot or overmixing}
\label{subsec:Overshoot}

\emph{Overshooting} (e.g.,~\citep{Zahn1991A&A...252..179Z}) or \emph{overmixing} (e.g.,~\citep{Woo2001AJ....122.1602W}) occurs when fluid motions beyond the convective boundary transport elements but not heat and thereby do not alter the temperature gradient. 
Two prescriptions for convective overshooting are common in 1D stellar evolution codes. 

\emph{Exponential diffusive overshoot} \citep{Freytag1996A&A...313..497F,Herwig2000A&A...360..952H}, see Figure~\ref{fig:CBM_1D} (a), is a 1D parameterization of results from 2D hydrodynamical simulations of surface convection zones in solar-type stars, main-sequence A-type stars, and cool DA white dwarfs. 
It is used by the stellar evolution code \texttt{GARSTEC} \citep{Weiss2008Ap&SS.316...99W}, and used to be the default overshoot option in \texttt{MESA} \cite{Paxton2011ApJS..192....3P,Paxton2013ApJS..208....4P,Paxton2015ApJS..220...15P,Paxton2018ApJS..234...34P,Paxton2019ApJS..243...10P}.
Exponential diffusive overshoot is a mixing efficiency which decreases exponentially with distance from the convective boundary,
\begin{equation}
    D_{\rm CBM} (r) = D_0 \exp \left[\frac{-2 \left( r - r_0\right)}{f_{\rm ov} H_{\rm p, 0}} \right]
    \qquad\mathrm{with}\qquad 
    \nabla_T = \nabla_{\rm rad}.
    \label{Eq:exp_ov}
\end{equation}

In this formalism, the free parameter $f_\text{ov}$ determines what fraction of a pressure scale height corresponds to the the $e$-folding length scale of the mixing efficiency, and thereby indirectly sets the extent of the CBM region; CBM models based on the scale height have been used as long as CBM has been considered \citep{roxburgh1965,Prather1974ApJ...193..109P}.
Here, $r$ is the radial coordinate and $H_p$ is the pressure scale height.
The convective boundary occurs at $r_{\rm{cc}}$ and has scale height $H_{p,\rm{cc}}$, but MLT assumes that the convective velocity and mixing are both zero at $r_{\rm cc}$.
As a result, exponential diffusive overshoot is calibrated at $r_0 = r_{\rm cc} - f_0 H_{p,\rm{cc}}$, where $f_0$ is usually fixed to a value between 0.002-0.005.
At $r_0$, the convective mixing efficiency is $D_0 = D_\text{mix} (r_0)$ and the pressure scale height is $H_{p,0}$.
The mixing efficiency follows the MLT value for $r < r_0$, and follows Eqn.~\ref{Eq:exp_ov} for $r \geq r_0$.
\footnote{To account for the step $f_0 H_\text{p,cc}$ taken inside of the convective core, one would usually add $f_0$ to the overshooting parameter. As an example, in \texttt{MESA} one would use \texttt{overshoot\_f}$= f_\text{ov} + f_0$, where \texttt{overshoot\_f} is the name of the overshoot parameter in \texttt{MESA}.} 

\emph{Step overshoot} provides a simpler mixing profile in the CBM region (see Figure~\ref{fig:CBM_1D}, b),
\begin{equation}
    D_{\rm CBM} (r) = D_0 \qquad \text{for} \ \ r_0 \leq r \leq r_\text{ov},  
    \qquad\mathrm{with}\qquad
    \nabla_T = \nabla_{\rm rad}.
\end{equation}
Here the free parameter $\alpha_{\rm{ov}}$ determines the extent $r_\text{ov} = r_0 + (\alpha_\text{ov} + f_0 ) H_\text{p, 0}$ of the overshoot region which is characterized by a constant mixing efficiency $D_0$. 
This CBM formalism is adopted in the stellar evolution codes \texttt{DSEP} \citep{Dotter2007AJ....134..376D,Dotter2008ApJS..178...89D}, 
\texttt{BaSTI} \citep{Cassisi1997MNRAS.285..593C,Salaris1998MNRAS.298..166S,Pietrinferni2004ApJ...612..168P,Pietrinferni2006ApJ...642..797P,Cordier2007AJ....133..468C,Percival2009ApJ...690..427P,Pietrinferni2009ApJ...697..275P},
\texttt{TGEC} \citep{HuiBonHoa2008Ap&SS.316...55H,Theado2012A&A...546A.100T}, and
\texttt{YREC} \citep{Demarque2008Ap&SS.316...31D}, and is available in \texttt{MESA}.

\subsection{Convective penetration}
\label{subsec:1d_conv_pen}

\emph{Convective Penetration} occurs when convection mixes chemicals and entropy beyond the convective boundary \citep{Zahn1991A&A...252..179Z}, see Figure~\ref{fig:CBM_1D} (c). 
Convective penetration is identical to step overshoot, except the adiabatic temperature gradient is adopted in the CBM region:
\begin{equation}
    D_{\rm CBM} (r) = D_0 \qquad \text{for} \ \ r_0 \leq r \leq r_\text{pen},  
    \qquad\mathrm{with}\qquad
    \nabla_T = \nabla_{\rm ad}.
\end{equation}
Here $r_\text{pen} = r_0 + (\alpha_\text{pen} + f_0 ) H_\text{p, 0}$, and $\alpha_\text{pen}$ is the free CBM parameter. The convective penetration formalism is adopted in the stellar evolution code \texttt{GENEC} \citep{Eggenberger2008Ap&SS.316...43E}, and a similar option is available for the \texttt{YREC} code. 
We caution that the names ``step overshoot'' and ``convective penetration'' are often used interchangeably in the literature.
We distinguish between the two based on the  temperature gradient in the CBM region (e.g.,~\citep{Zahn1991A&A...252..179Z,Viallet2015A&A...580A..61V}).

\subsection{Extended convective penetration}
\label{subsec:ext_conv_pen} 

\textit{Extended convective penetration} \citep{Michielsen2019A&A...628A..76M,Michielsen2021A&A...650A.175M} combines convective penetration and diffusive exponential overshooting, see Figure~\ref{fig:CBM_1D} (d).
In this formalism, the convective boundary mixing region has two components.
The convection zone is adjacent to a convective penetrative region with constant mixing and an adiabatic temperature gradient. 
Further from the convective boundary, the mixing exponentially decays and the temperature gradient gradually transitions from $\gradad$ to $\gradrad$. The exact mixing coefficients are

\begin{align}
D_\text{CBM}(r) = D_\text{pen} \qquad &\text{for} \ \ r_0 \leq r \leq r_\text{pen}\nonumber\\
D_\text{CBM}(r) = D_\text{pen} \exp\left[\frac{-2 (r - r_\text{pen})}{f_\text{pen} H_\text{p,pen}}\right] \qquad &\text{for} \ \ r_\text{pen} < r \leq r_\text{CBM},
	\label{Eq:CBM_ConvPenExp}
\end{align}

where $H_\text{p,pen} = H_\text{p}(r_\text{pen})$ and $r_\text{CBM}$ is the radius at the outer boundary of the CBM region. The thermal stratification is  \cite{Michielsen2019A&A...628A..76M}

\begin{align}
\nabla_\text{T} = \nabla_\text{ad} \qquad &\text{for} \ \ r_0 \leq r \leq r_\text{pen}\nonumber\\
\nabla_\text{T} =  f^{t}(r) \nabla_\text{ad} + \left[1-f^{t}(r)\right]\nabla_\text{rad}\qquad &\text{for} \ \ r_\text{pen} < r \leq r_\text{CBM},
	\label{Eq:CBM_ConvPenExp}
\end{align}

There are two free parameters: $\alpha_\text{pen}$ and $f_{\rm pen}$. 
$f^{t}(r)$ is a radial profile which varies from one in the convection zone to zero in the stable radiative envelope.
$f^{t}(r)$ has been prescribed in two ways in the literature. 
Its first implementation was based on the amount of mass in $r_\text{pen} < r \leq r_\text{CBM}$ \citep{Michielsen2019A&A...628A..76M}. 
Another implementation define it as $f^{t}(r) = \left[\log Pe(r) + 2\right]/4$ for values of $10^{-2} < Pe < 10^{2}$, where the P{\'e}clet number $Pe$ \citep{Michielsen2021A&A...650A.175M}, which is the ratio between the time scales of the radiative and advective heat transport and can be estimate from e.g., MLT velocities. 

To our knowledge, extended convective penetration is not currently a standard option in any stellar evolution codes, but it has been studied using a modified version of \texttt{MESA} \citep{Michielsen2019A&A...628A..76M,Michielsen2021A&A...650A.175M}. 
However, the option of changing the temperature gradient within the CBM region is available in existing codes. 
For example, the \texttt{ASTEC} code \citep{ChristensenDalsgaard2008Ap&SS.316...13C} uses the same mixing profile $D_{\rm CBM}$ as the step overshoot and convective penetration schemes wherein mixing efficiency is considered constant for a certain distance $\alpha_\text{CBM} H_p$ beyond the convective boundary, but \texttt{ASTEC} can smoothly vary the temperature gradient within this region \citep{ChristensenDalsgaard2008Ap&SS.316...13C,Monteiro1994A&A...283..247M}.


\subsection{Limiting the extent of the CBM region}
\label{subsec:cbm_limits}

The diffusive overshoot, convective penetration, and extended convective penetration prescriptions listed above all rely on a free CBM parameter multiplied by the pressure scale height to define the extent of the CBM region. 
As a result, small convective cores (with $r_\text{cc} \rightarrow 0$ and $H_\text{p} \rightarrow \infty$) can produce unphysically large CBM regions.
This problem primarily arises in lower mass stars that start off with small convective cores on the main sequence.
To circumvent this issue, some stellar evolution codes implement a mass-dependent CBM parameter which is zero at low stellar mass, constant at high mass, and smoothly increases at intermediate mass.
Such an approach is partially validated by observational evidence of a relationship between CBM mixing parameters and stellar mass, but the observational evidence is ambiguous, see Section~\ref{sec:observations}. 
Mass-dependent CBM parameters were used to compute the \texttt{YREC} Y$^2$ isochrones \cite{Demarque2004ApJS..155..667D}, the YaPSI isochrones \cite{Spada2017ApJ...838..161S}, a set of isochrones computed with \texttt{GARSTEC} and used to fit the open cluster M67 \cite{Magic2010ApJ...718.1378M}, and a grid of stellar models with derived internal structure constants computed with \texttt{MESA} \cite{Claret2019A&A...628A..29C}. 

Various alternative approaches limit the size of CBM regions based on the size of the convection zone. 
The \texttt{ASTEC} and \texttt{Cesam2k} codes enforce that the size of the CBM region is $\beta_\text{CBM}\, \times \text{min} (r_{\rm cc}, H_{\rm p, cc})$ \citep{ChristensenDalsgaard2008Ap&SS.316...13C,Morel2008Ap&SS.316...61M}. The \texttt{YREC} code defines the actual radius of the CBM region as $r_\text{CBM} = \beta_\text{CBM} / (H_\text{p}^{-1} + r_\text{cc}^{-1})$ which simplifies to $r_\text{CBM} = \beta_\text{CBM} H_\text{p}$ for large convective cores \cite{Viani2020ApJ...904...22V}. The \texttt{GARSTEC} code uses a modified pressure scale height $\widetilde{H_\text{p}} = H_\text{p} \times \text{min} \left[1, \left( r_\text{cc}/H_\text{p}\right)^2 \right]$ \cite{Weiss2008Ap&SS.316...99W,Magic2010ApJ...718.1378M}, and as a default \texttt{MESA} uses $\beta_\text{CBM} \times \text{min}\left(H_\text{p}, r_\text{cc}/\alpha_\text{MLT}\right)$ where $\alpha_\text{MLT}$ is the mixing length parameter. 
Here we use $\beta_\text{CBM}$ to collectively refer to the free parameter assumed for a given CBM prescription, so it could be e.g., $f_\text{ov}$ or $\alpha_\text{ov}$.
A result of these constraints is that the input $\beta_\text{CBM}$ parameter in the code can be different from the effective $\beta_\text{CBM}$ used to set the size of the CBM region (see Ref.~\citep{Viani2020ApJ...904...22V}, Figures 2 and 5 for examples of this). 
A further problem is that inconsistent implementations between different codes produce CBM regions with different sizes even if the same $\alpha_\text{CBM}$ parameter and CBM prescription are nominally employed.
These subtle differences impede direct comparisons between results obtained using different codes when the sizes of the convective cores are small as is the case for stars around $1.2$\,M$_\odot$ \cite{Deheuvels2019BSRSL..88...84D}.


\subsection{Comparing different CBM parameters}
\label{subsec:comp_params}

In order to make comparisons between results obtained using different CBM prescriptions (e.g., exponential vs.~step overshooting), a robust conversion between their input parameters must be established.
Such conversions can be achieved in two ways. 

The first is to compare observations to models generated using different CBM prescriptions and find the CBM parameters which best reproduce the observed diagnostics. 
Such comparisons have previously been achieved through, e.g., the asteroseismic modeling of the slowly pulsating B star KIC~7760680 where comparisons between results using exponential diffusive and step overshoot suggested a relation of $\alpha_\text{ov} /f_\text{ov} = 13.33$ for this $3.25$\,M$_\odot$ star \cite{Moravveji2016ApJ...823..130M}, see also Section~\ref{subsec:asteroseismology}. 
Similarly, comparisons have been made using detached, double-lined, eclipsing binary stars (see also Section~\ref{subsec:mass_discrepancy}). 
As an example, the study of 29 such systems with component masses between 1.2 and 4.4\,M$_\odot$ revealed a relation of $\alpha_\text{pen} /f_\text{ov} = 11.36 \pm 0.22$ between models using convective penetration and exponential diffusive overshoot \citep{Claret2017ApJ...849...18C}. 
The comparison suggested that there is a slight dependence of this ratio on surface gravity $\log g$, metallicity $Z$, mass $M$, or effective temperature $T_\text{eff}$. Splitting the sample in two groups depending on either the effective temperature or surface gravity resulted in $\alpha_\text{pen} /f_\text{ov} = 10.50 \pm 0.25$ for cooler giant stars and $\alpha_\text{pen} /f_\text{ov} = 11.71 \pm 0.27$ for hotter dwarf stars \citep{Claret2017ApJ...849...18C}. 
A similar study of 12 binary systems with component masses between 4.58 and 17.07\,M$_\odot$ likewise suggest a conversion factor between $f_\text{ov}$ and $\alpha_\text{pen}$ larger than 10 \cite{Tkachenko2020A&A...637A..60T}\footnote{We note that the authors of both of these studies of detached double-lined eclipsing binary systems  \cite{Claret2017ApJ...849...18C,Tkachenko2020A&A...637A..60T} use $\alpha_\text{ov}$ in their notation, but are actually assuming an adiabatic temperature gradient in the CBM region. 
In other words, while they talk about a step-based overshooting using the free parameter $\alpha_\text{ov}$, they are in fact referring to convective penetration.}.

Another method for deriving conversions between different CBM parameters is to directly compare models which use different CBM prescriptions. 
\citet{Magic2010ApJ...718.1378M} suggested a conversion factor of $\alpha_\text{ov} /f_\text{ov} \approx 11$ using models with masses between 2 and 6\,M$_\odot$. 
\citet{Noels2010Ap&SS.328..227N} compared two evolutionary tracks of a 10\,M$_\odot$ star using either step overshoot or convective penetration, finding that an $\alpha_\text{pen} = 0.175$ achieved a similar result to the step overshoot case with $\alpha_\text{ov} = 0.2$, i.e. $\alpha_\text{ov}/\alpha_\text{pen} = 1.14$.

\begin{figure}[t]
\includegraphics[width=\linewidth]{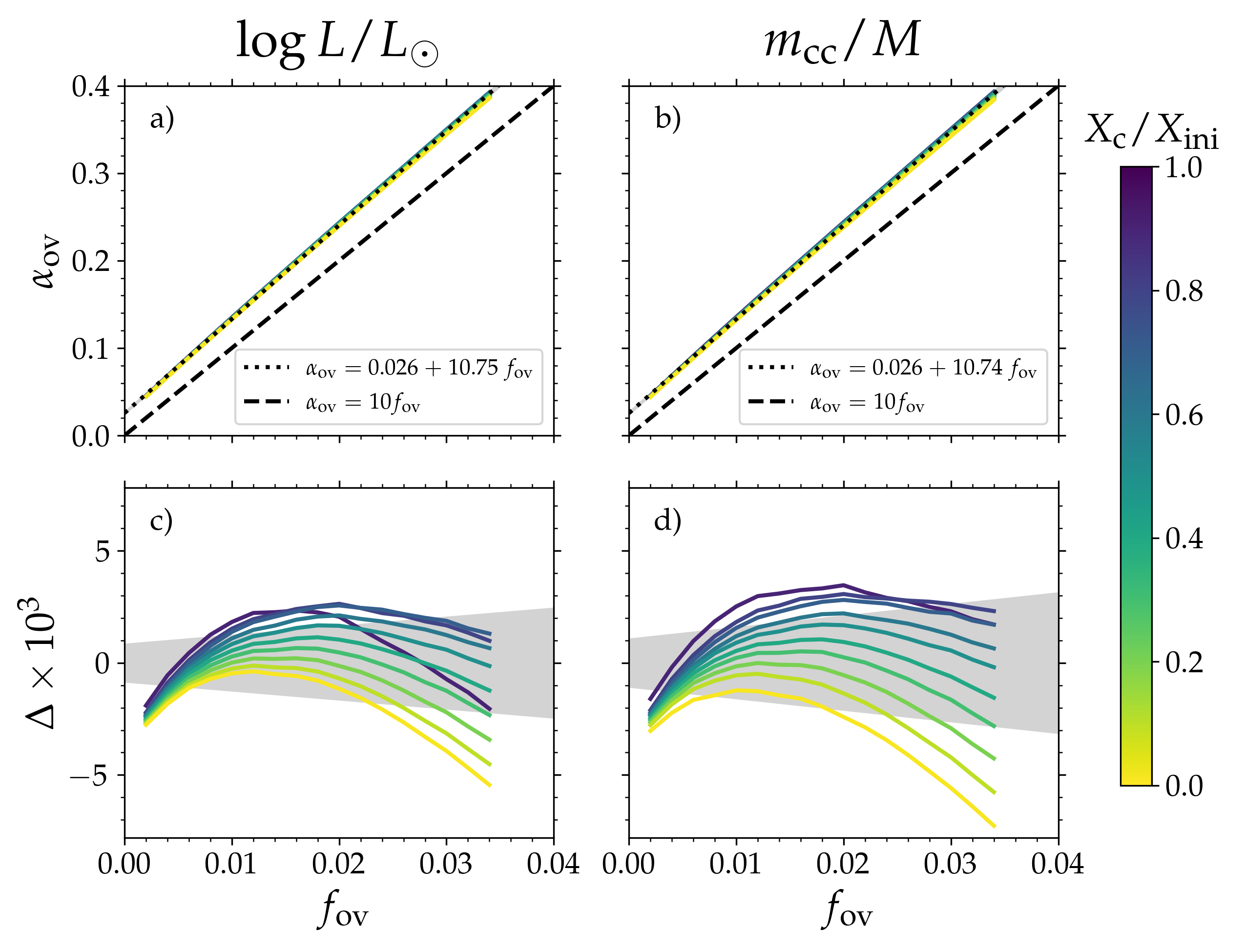}
\caption{Derived correlations between $\alpha_\text{ov}$ and $f_\mathrm{ov}$ for a 10\,M$_\odot$ star at different main-sequence ages given as $X_\text{c}/X_\text{ini}$. (\textbf{a}) The colored curves show the relation between $\alpha_\text{ov}$ and $f_\text{ov}$ required to obtain $\log L_1 \left( X_\text{c}/X_\text{ini}, f_\text{ov} \right) = \log L_2 \left( X_\text{c}/X_\text{ini}, \alpha_\text{ov} \right)$. The black dashed curve shows the expected trend assuming $\alpha_\text{ov} = 10 f_\text{ov}$, while the black dotted lines shows the combined linear fit to the colored lines. (\textbf{b}) Same as (\textbf{a}) but fulfilling $m_{cc,1} \left( X_\text{c}/X_\text{ini}, f_\text{ov} \right) / M = m_{cc,2} \left( X_\text{c}/X_\text{ini}, \alpha_\text{ov} \right) / M$. (\textbf{c}) Differences between the derived $\alpha_\text{ov}$ versus $f_\mathrm{ov}$ relation for a given $X_\text{c}/X_\text{ini}$ and the linear fit shown by the dotted black line in panel \textbf{a}. The grey shaded region gives the $3\,\sigma_\text{std}$ uncertainty region of the linear fit. (\textbf{d}) Same as (\textbf{c}) but for the comparison between the $\alpha_\text{ov}$ and $f_\text{ov}$ parameters needed to get the same convective core masses. Figure made by the authors using \texttt{MESA} models. \texttt{MESA} inlists and data used to generate the figure are available on Zenodo \citep{supp}.
\label{fig:fov_vs_alphaov}}
\end{figure}   

Here we provide a comparison between models computed with the stellar structure and evolution code \texttt{MESA} v22.05.1 for a 2\,M$_\odot$ and 10\,M$_\odot$ star assuming 1) exponential diffusive overshoot, and 2) diffusive step overshoot. Using these models, we investigate what $\alpha_\text{ov}$ parameter is required to reproduce the same luminosity and convective core mass ($m_{cc}$, the mass coordinate of the Schwarzschild boundary of the convection zone) of the star with exponential diffusive overshoot at a fixed value of $f_\text{ov}$. These comparisons are carried out at fixed main-sequence age ($X_\text{c}/X_\text{ini}$) and stellar mass.
In other words, we look for solutions to the relations $\log L_1 \left( X_\text{c}/X_\text{ini}, f_\text{ov} \right) = \log L_2 \left( X_\text{c}/X_\text{ini}, \alpha_\text{ov} \right)$ and $m_{cc,1} \left( X_\text{c}/X_\text{ini}, f_\text{ov} \right) / M = m_{cc,2} \left( X_\text{c}/X_\text{ini}, \alpha_\text{ov} \right) / M$. Here $X_\text{c}$ is the current core hydrogen mass fraction and $X_\text{ini}$ is the initial hydrogen mass fraction. An example of these solutions is shown for the 10\,M$_\odot$ star in Figure~\ref{fig:fov_vs_alphaov}. Panels (\textbf{a}) and (\textbf{b}) show the derived relations for the luminosity and convective core mass, respectively, at different main-sequence ages indicated by the color of the lines. The black dashed curve shows the expected trend assuming the standard ``rule-of-thumb'' $\alpha_\text{ov} \approx 10 f_\text{ov}$, whereas the black dotted line shows the linear fit to the derived relations. The differences between the linear fits and relations derived at different ages are shown in panel (\textbf{c}) and (\textbf{d}), where the grey shaded region gives the $3\sigma_\text{std}$ uncertainty regions for the relations. As seen in the figure, the relations are not strictly linear and also show a small dependence on the chosen $X_\text{c}$ in the current core hydrogen mass fraction and $X_\text{ini}$ value.


In summary, we find a general relation of the form
\begin{equation}
\alpha_{\rm ov} = A + B f_{\rm ov}.
\end{equation}
For the 2 $M_{\odot}$ star, we find $A = (0.042 \pm 0.004)$ and $B = (14.11 \pm 0.25)$ when fitting for luminosity and $A = (0.036 \pm 0.01)$ and $B = (14.05 \pm 0.71)$ when fitting for core mass.
For the 10 $M_{\odot}$ star, we find $A = (0.0256 \pm 0.0008)$ and $B = (10.75 \pm 0.04)$ when fitting for luminosity and $A = (0.0256 \pm 0.001)$ and $B = (10.74 \pm 0.05)$ when fitting for core mass.
The reported errors are the $3\sigma_\text{std}$ errors.

We note three important observations for the four relations given above. The first is that relations between $\alpha_\text{ov}$ and $f_\text{ov}$ for a given mass are same within the errors independently of whether they are derived based on the luminosity or the convective core mass. The second observation is that the errors on the parameters for the relations for the 2\,M$_\odot$ star are larger than for the 10\,M$_\odot$, implying a stronger age dependence on the relations for the 2\,M$_\odot$ star. Finally, the slopes ($B$) of the relations are steeper for the 2\,M$_\odot$ star than the 10\,M$_\odot$ one and cannot be reconciled within the $3\sigma_\text{std}$ errors, showing that the exact conversions to use are also mass dependent.
We emphasize that making direct comparisons between absolute values of different CBM parameters is non-trivial. Therefore, when studying CBM and the extent of the CBM region, carrying out an ensemble study of a group of stars using the same stellar evolution code and the same CBM prescription is recommended.

\subsection{1D models not covered in this review}

    A full discussion of physically-motivated 1D models of CBM is beyond the scope of this review.
    Techniques not discussed here include CBM models based on linear or fundamental mode analysis \citep{unno1957, bohm1963, saslaw_schwarzschild_1965}, nonlinear modal expansion \citep{zahn_etal_1982}, models of overshooting bubbles based on local MLT \citep{roxburgh1965, straus_etal_1976}, nonlocal MLT models \citep{spiegel_1963, shaviv_salpeter_1973, nordlund1974, maeder1975, cogan1975, ulrich1976, langer1986}, non-MLT multiscale models \citep{marcus_etal_1983}, ``turbulent convection models'' (e.g.~\citep{zhang_li_2012,zhang2016}), Canuto's stellar mixing model (e.g.,~\citep{canuto2011}), and non-local ``turbulent kinetic energy models'' (e.g.,~\citep{kupka_montgomery_2002,kupka_etal_2022}).
    We briefly also note that there exist models which aim to characterize overshooting convective motions in the optically-thin atmosphere of stars like the Sun \citep{unno1957,chen1974}, reviewed briefly in \citet{nordlund1974}, but we focus here on convection confined to optically thick portions of the star.

    In this section, we have focused on the most frequently-used techniques in the stellar modeling literature; we note that these MLT-based implementations may not necessarily be logically consistent \citep{renzini1987}, but their ease of implementation and use has made them widespread in the stellar structure literature.
    Other local-MLT-based CBM prescriptions or profiles such as diffusive Gumbel overshooting \cite{Pratt2017A&A...604A.125P,Augustson2019ApJ...874...83A} and diffusive double exponential overshoot \cite{Herwig2007ASPC..378...43H} have been proposed, but a full discussion of them is likewise beyond the scope of this review.

\section{3D Hydrodynamical Perspectives of Convective Boundary Mixing}
\label{sec:simulations}
We now describe convective boundary mixing from its fluid dynamical roots.
Many simulations have examined a convective layer interacting with a stable layer in a variety of natural contexts (e.g., stellar envelope convection, core convection, and even atmospheric convection).
In order to paint the most complete picture of convective boundary mixing, we will discuss the results of these studies from a perspective that is impartial to the motivation or setup.

Hydrodynamical CBM simulations often employ simplified equation sets (e.g., the Boussinesq \citep{spiegel_veronis_1960} or Anelastic  \citep{lantz1992,braginsky_roberts_1995,lantz_fan_1999,jones_etal_2009} approximations\footnote{The Anelastic approximation models low Mach number flows and assumes that Eqn.~\ref{eqn:continuity} reduces to $\nabla\cdot(\rho_0\vec{u}) = 0$ where $\rho_0$ is the ``background'' density. The Boussinesq approximation goes one step further and assumes incompressibility, or that $\rho_0$ is constant everywhere so that Eqn.~\ref{eqn:continuity} becomes $\nabla\cdot\vec{u} = 0$; under the Boussinesq approximation, small density perturbations are allowed to exist in the buoyancy term in the momentum equation.}).
For generality, we will use the equation formulation most applicable to stellar interiors, the Fully Compressible Navier-Stokes equations
(Ref.~\citep{landau_lifshitz_book}, $\S$15 and $\S$49), which are
\begin{align}
    \partial_t \rho + \grad\cdot(\rho\vec{u}) = 0,
    \label{eqn:continuity} \\
    \partial_t \vec{u} + \vec{u}\cdot\grad\vec{u} = -\frac{1}{\rho}\grad P + \vec{g} + \frac{1}{\rho}\grad\cdot\vec{\overline{\Pi}},
    \label{eqn:momentum} \\
    \partial_t s + \vec{u}\cdot\grad s = \frac{1}{\rho T} (\grad\cdot(k \grad T) + \rho \epsilon + \Phi),
    \label{eqn:energy}
\end{align}
where $\rho$ is the density, $\vec{u}$ is the velocity, $T$ is the temperature, $s$ is the specific entropy, $\vec{g}$ is the gravitational acceleration vector, $k$ is the radiative conductivity, and $\epsilon$ is the specific energy production rate (erg g$^{-1}$\,s$^{-1}$) from nuclear burning.
The viscous stress tensor, viscous heating, and rate-of-strain tensor are respectively defined
\begin{gather}
    \Pi_{ij} = 2 \rho \nu \left[e_{ij} - \frac{1}{3}(\grad\cdot\vec{u})\delta_{ij}\right],
    \label{eqn:stress_tensor}
    \\
    \Phi = 2\rho\nu\left[e_{ij}e_{ij} - \frac{1}{3}(\grad\cdot\vec{u})^2\right]
    = 2\rho\nu\left[\mathrm{Tr}(\overline{\vec{e}}\cdot\overline{\vec{e}}) - \frac{1}{3}(\grad\cdot\vec{u})^2\right],
    \label{eqn:visc_heat}
    \\
    e_{ij} = \frac{1}{2}(\grad \vec{u} + [\grad \vec{u}]^T).
    \label{eqn:strain_tensor}
\end{gather}
where $\nu$ is the viscous diffusivity (kinematic viscosity).
Stars are composed of magnetized plasmas and thus magnetohydrodynamic effects should in general be accounted for, but for simplicity we will restrict our discussion to the hydrodynamic case in this review.

Arguments about CBM processes are often rooted in energetics.
The kinetic energy equation is obtained by dotting Eqn.~\ref{eqn:momentum} with $\rho\vec{u}$ and applying Eqn.~\ref{eqn:continuity} to retrieve
\begin{equation}
    \partial_t (\mathcal{KE} + \mathcal{PE}) + \grad\cdot(\vec{u}(\mathcal{KE} + \mathcal{PE} + P) + \vec{u}\cdot\vec{\overline{\Pi}}) = P\grad\cdot\vec{u} - \Phi.
    \label{eqn:ke_equation}
\end{equation}
Here, the kinetic energy is $\mathcal{KE} = \rho |\vec{u}|^2 / 2$ and the potential energy is $\mathcal{PE} = \rho \phi$, and we have assumed time invariance $\partial_t \phi = 0$ of the gravitational potential $\phi$ (defined from $\vec{g} = -\grad\phi$).
Eqn.~\ref{eqn:ke_equation} is written in conservation form, with the time derivative and divergence of energy fluxes on the left-hand side and the sources and sinks of energy on the right-hand side.
An entropy equation is obtained by multiplying Eqn.~\ref{eqn:energy} by $\rho$ and applying Eqn.~\ref{eqn:continuity},
\begin{equation}
    \partial_t(\rho s) + \grad\cdot(\vec{u}\rho s) = \frac{1}{T}\left(\grad\cdot(k \grad T) + \rho \epsilon + \Phi\right).
    \label{eqn:entropy_eqn}
\end{equation}
We note that we could have instead multiplied by $\rho T$ to obtain the internal energy equation, but that would generally return the same constraints as the kinetic energy equation, so a different thermal energy constraint is needed.

We next take a volume integral (denoted by angle brackets $\langle\rangle$) of Eqns.~\ref{eqn:ke_equation} \& \ref{eqn:entropy_eqn} over the full convection zone and any important CBM region.
We apply the divergence theorem to the flux terms and assume that the volume we are examining is bounded by regions where $\vec{u} \approx 0$, so the integral of the fluxes can be neglected.
We are left with
\begin{align}
    \partial_t (\angles{\mathcal{KE}} + \angles{\mathcal{PE}}) = \angles{P\grad\cdot\vec{u}} - \angles{\Phi},
    \label{eqn:integral_template} \\
    \partial_t {\angles{\rho s}} = \angles{\frac{1}{T}\left[\grad\cdot(k\grad T) + \rho \epsilon\right]} + \angles{\frac{1}{T}\Phi}.
    \label{eqn:therm_integral_template}
\end{align}
Eqn.~\ref{eqn:integral_template} states that the evolution of the total (kinetic and potential) energy of the convection zone is determined by the fraction of $P \text{d}V$ work ($\angles{P\grad\cdot\vec{u}}$) that is not consumed by dissipative processes ($\angles{\Phi}$) on small scales.
Eqn.~\ref{eqn:therm_integral_template} forms the basis for deriving thermal scaling laws for convective regions \citep{grossman_lohse_2000,jones_etal_2022} and also serves as a basis for the integral constraint of \citet{roxburgh_1989}.

\begin{figure}
    \centering
    \includegraphics[width=\textwidth]{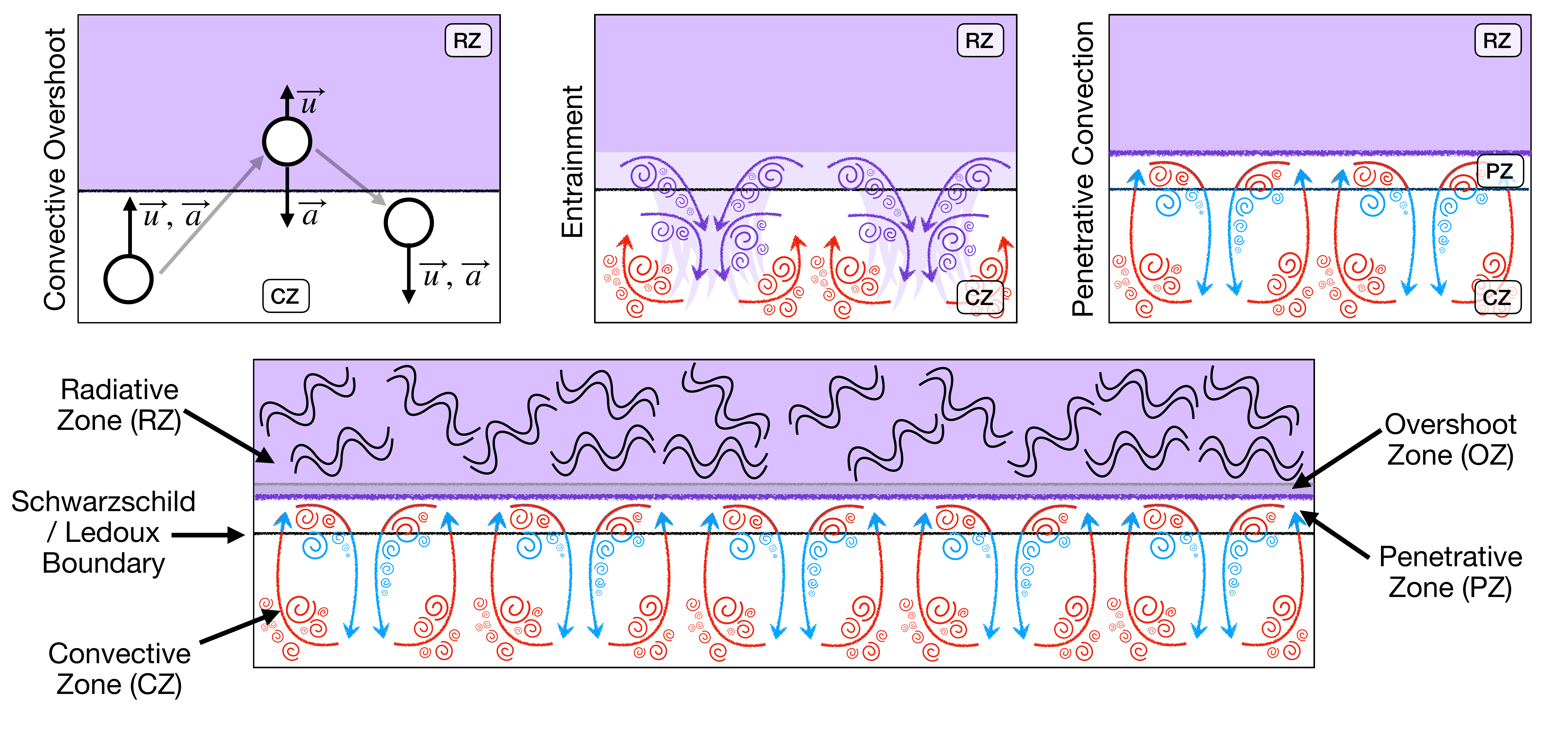}
    \caption{
    Three CBM processes are shown schematically in the top row. 
    White fluid represents the well-mixed CZ, while purple fluid is the stable RZ. 
    (Left) Convective overshoot (Section \ref{subsec:3d_overshoot}) occurs when a fluid parcel from the CZ crosses into the RZ; a positive entropy gradient in the RZ accelerates the parcel back toward the CZ. 
    (Middle) Shear flows and overshooting motions drag RZ fluid into the CZ in a process called entrainment (Section \ref{subsec:3d_entrainment}). 
    (Right) Convective penetration occurs when convection maintains a well-mixed region of fluid beyond the Schwarzschild boundary (Section \ref{subsec:3d_penetration}). 
    The bottom panel shows the structure of a statistically-stationary convective boundary, which resembles the 1D  ``extended convective penetration'' prescription (Section \ref{subsec:ext_conv_pen}).
    This figure was originally published online under a CC BY license in \citet{anders_rnaas}. }
    \label{fig:processes_structure}
\end{figure}

We will use this energetics framework to describe three \emph{processes} that can occur in hydrodynamical CBM.
These processes are depicted in Figure~\ref{fig:processes_structure} and parallels can be drawn between these processes and the prescriptions in Section~\ref{sec:parameterizations}.
The first process is a small-scale conversion of convective kinetic energy into potential energy beyond the boundary, which is referred to as ``convective overshoot.''
The second is a process wherein either kinetic energy or buoyant work are used to increase the potential energy of the convection zone, referred to as ``entrainment.''
The third process occurs in a statistically stationary state where $\partial_t(\angles{\mathcal{KE}} + \angles{\mathcal{PE}}) = 0$, and a balance between work producing energy and dissipation is achieved; this process is called ``convective penetration.''
We note that there is a great deal of degeneracy in the literature studying these processes, and these terms (in particular ``overshoot'' and ``penetration'') are often used interchangeably; note that when we use these terms in this review we are referring to distinct processes.
As in the bottom panel of Figure~\ref{fig:processes_structure}, we will assume that the stellar structure consists of a convection zone (CZ, $\grad \approx \gradad < \gradrad$) and a radiative zone (RZ, $\grad \approx \gradrad < \gradad$), and that the CBM region consists of both a penetrative zone (PZ, $\grad \approx \gradad > \gradrad$) and an overshoot zone (OZ, $\grad \approx \gradrad < \gradad$).
We note that the convection zone itself could also have additional structure (e.g., ``Deardorff zones'', see Ref.~\citep{kapyla_etal_2017}), but we do not include that level of detail here.

\subsection{Nondimensional Fluid Parameters}
\label{subsec:fluid_params}
Many hydrodynamical studies of convective boundary mixing processes seek a description of how a CBM length scale or rate varies as a nondimensional fluid parameter is changed.
Commonly measured nondimensional numbers are the bulk Richardson number, $\Ri$, and the stiffness or relative stability, $\mS$.
These parameters compare the buoyancy stability of the RZ to how unstable or vigorous the convection is in the CZ.

The Richardson number was first examined in the astrophysical context by \citet{meakin_arnett_2007} and is typically defined
\begin{equation}
    \Ri = \frac{\Delta b L}{\sigma^2},
    \qquad\mathrm{with}\qquad
    \Delta b = \int_{r_1}^{r_2} N^2\, dr,
\end{equation}
where $\sigma$ is the root-mean-square turbulent velocity in the convective region near the interface, $N^2$ is the \brunt~frequency, $L$ is a typical length scale for turbulent motions, and the convective interface is assumed to span some radial extent ranging from $r_1$ to $r_2$.
There is degeneracy in how $r_1$, $r_2$, and $L$ are defined in the literature.

An alternative approach is to measure the ``stiffness'' or ``relative stability'' $\mS$ of the radiative-convective interface.
This has historically been defined in two ways.
Early simulations defined a structure-based $\mS$ \cite{hurlburt_etal_1994}\footnote{$\mS_{\rm struct}$ was often defined in terms of polytropic indices, thus we use $\approx$ instead of $=$ in our definition here.},
\begin{equation}
    \mS_{\rm struct} \approx \frac{|\gradrad - \gradad|_{\rm RZ}}{|\gradrad - \gradad|_{\rm CZ}},
\end{equation}
where the logarithmic temperature gradients are defined in Sct.~\ref{subsec:convboundaries}.
This definition is useful in simulations where convection is driven by an unstable temperature gradient which achieves $\grad = \gradrad$ by e.g., an enforced boundary condition, but it is less useful in describing convection in the cores of massive stars where $\grad \approx \gradad \ll \gradrad$.
Recently, a dynamical definition of the stiffness has been favored by some authors \citep{lecoanet_etal_2016,couston_etal_2017,anders_etal_2022a},
\begin{equation}
    \mS_{\rm dyn} = \frac{N_{\rm RZ}^2}{\omega_{\rm conv}^2},
\end{equation}
where $N_{\rm RZ}^2$ is the typical value of the \brunt~frequency in the stable radiative zone and $\omega_{\rm conv}^2 = [2\pi u_{\rm conv} / L]^2$ is the square angular convective frequency, where $u_{\rm conv}$ is the average turbulent convective velocity and $L$ is a typical convective length scale.
We then see that $\mS_{\rm dyn} \approx \Ri$, aside from the length scales which are used.
It is generally expected that stiff convective interfaces (with large values of $\mS$ or $\Ri$) should have very small CBM regions.
Stellar evolution models \citep{jermyn_etal_2022_atlas} and asteroseismic inferences \citep{aerts_etal_2021_stiffness} expect the stiffness value at the core boundaries of massive stars to be very large ($\mS_{\rm dyn} \sim 10^{6-8}$).

We also note that there is an explicit link between the Mach number $\mathrm{Ma}^2 = u_{\rm conv}^2 / c_s^2$ of convection and $\mS_{\rm dyn}$; knowledge of Ma in a convection zone therefore provides information about CBM.
Take $c_s^2 = P / \rho = g H_P$ to be the sound speed in a star in hydrostatic equilibrium, where $H_P$ is the pressure scale height and $g$ is the gravitational acceleration.
Neglecting composition gradients, and assuming $N^2 = (g/H_P) (\gradad - \gradrad)$ in the RZ  (Equation 6.18 of Ref.~\citep{kippenhahn_book}) and  $\omega_c^2 = [2\pi u_{\rm conv} / H_P]^2$ in the CZ, it can be shown that
\begin{equation}
    \mS_{\rm dyn} = \mathrm{Ma}^{-2} \frac{(\gradad - \gradrad)_{\rm RZ}}{(2\pi)^2}.
\end{equation}

\citet{anders_etal_2022a} recently introduced a new ``penetration parameter'' $\mathcal{P}$ to the zoo of parameters that describe CBM.
The extent of an adiabatic penetration zone is assumed to be determined by the magnitude of the negative buoyant work done within this zone \citep{roxburgh_1989,Zahn1991A&A...252..179Z}.
Therefore, a luminosity (or flux) based parameter can be defined \citep{anders_etal_2022a},
\begin{equation}
    \mathcal{P} = - \frac{(L_{\rm rad} - L_{\rm{ad}})_{\rm CZ}}{(L_{\rm rad} - L_{\rm ad})_{\rm RZ}},
\end{equation}
where the numerator (CZ) is averaged over some part of the convective zone and the denominator (RZ) is averaged over some part of the region that would be a radiative zone if not for convective penetration.
Here, $L_{\rm ad}$ is the radiative luminosity that would be carried if the stratification were adiabatic $\grad = \gradad$, and $L_{\rm rad}$ is the radiative luminosity that would be carried if the stratification were $\grad = \gradrad$; Since $(L_{\rm rad} - L_{\rm ad})_{\rm RZ} < 0$, $\mathcal{P} > 0$ always.
Large penetrative regions are expected when $\mathcal{P}$ is large.
There is an implicit link between the penetration parameter and the structural relative stability parameter, $\mathcal{P} \sim \mathcal{S}_{\rm struct}^{-1}$.
It therefore makes sense that simulations (e.g.,~\citep{brummell_etal_2002,rogers_glatzmaier_2005}) see negligible penetration when they use $\mathcal{S}_{\rm struct} \gg 1$.
Note that the dependence of convective penetration on $\mathcal{P}$ or $\mS_{\rm struct}$ is why we distinguish between $\mS_{\rm struct}$ and $\mS_{\rm dyn}$.
Many simulations are set up in such a manner that $\mS_{\rm struct} \sim \mS_{\rm dyn}$; however, stars can have $\mathcal{P} \sim \mS_{\rm struct} \sim 1$ and $\mathcal{S}_{\rm dyn} \gg 1$ simultaneously.

We finally note that studies dating back to \citet{Zahn1991A&A...252..179Z} \& \citet{hurlburt_etal_1994} ponder the importance of the P\'{e}clet number on CBM (e.g.,~\citep{brummell_etal_2002,browning_etal_2004,rempel2004,rogers_etal_2006,brun_etal_2011,Viallet2015A&A...580A..61V}).
The P\'{e}clet number measures the ratio of the thermal diffusion timescale to the convective overturn timescale,
\begin{equation}
    \mathrm{Pe} = \frac{\tau_{\rm therm}}{\tau_{\rm conv}} = \frac{u_{\rm conv} L}{\chi_{\rm rad}},
\end{equation}
where $\chi_{\rm rad}$ is the radiative diffusivity of the fluid.
The associated argument suggests that CBM regions have an adiabatic penetration zone where $\mathrm{Pe}$ is large as well as a ``thermal adjustment layer'' where $\mathrm{Pe} \sim 1$.
The size of the CBM region is expected to scale like $\mS_{\rm struct}^{-1}$ for the adiabatic penetration zone and like $\mS_{\rm struct}^{-1/4}$ for the thermal adjustment layer \citep{hurlburt_etal_1994}.
While these scalings were observed by early simulations (e.g.,~\citep{singh_etal_1998,saikia_etal_2000}), they have not appeared in more recent simulations \citep{brummell_etal_2002,rogers_glatzmaier_2005}.
Main-sequence convection is very turbulent with bulk $\mathrm{Pe} \gg 1$ \citep{jermyn_etal_2022_atlas}, so it is hard to imagine that a large thermal adjustment layer should appear at the radial location where convective flows have damped to the point of becoming laminar ($\mathrm{Pe} \sim 1$).

\subsection{Convective Overshoot}
\label{subsec:3d_overshoot}
\begin{figure}[t]
\includegraphics[width=\linewidth]{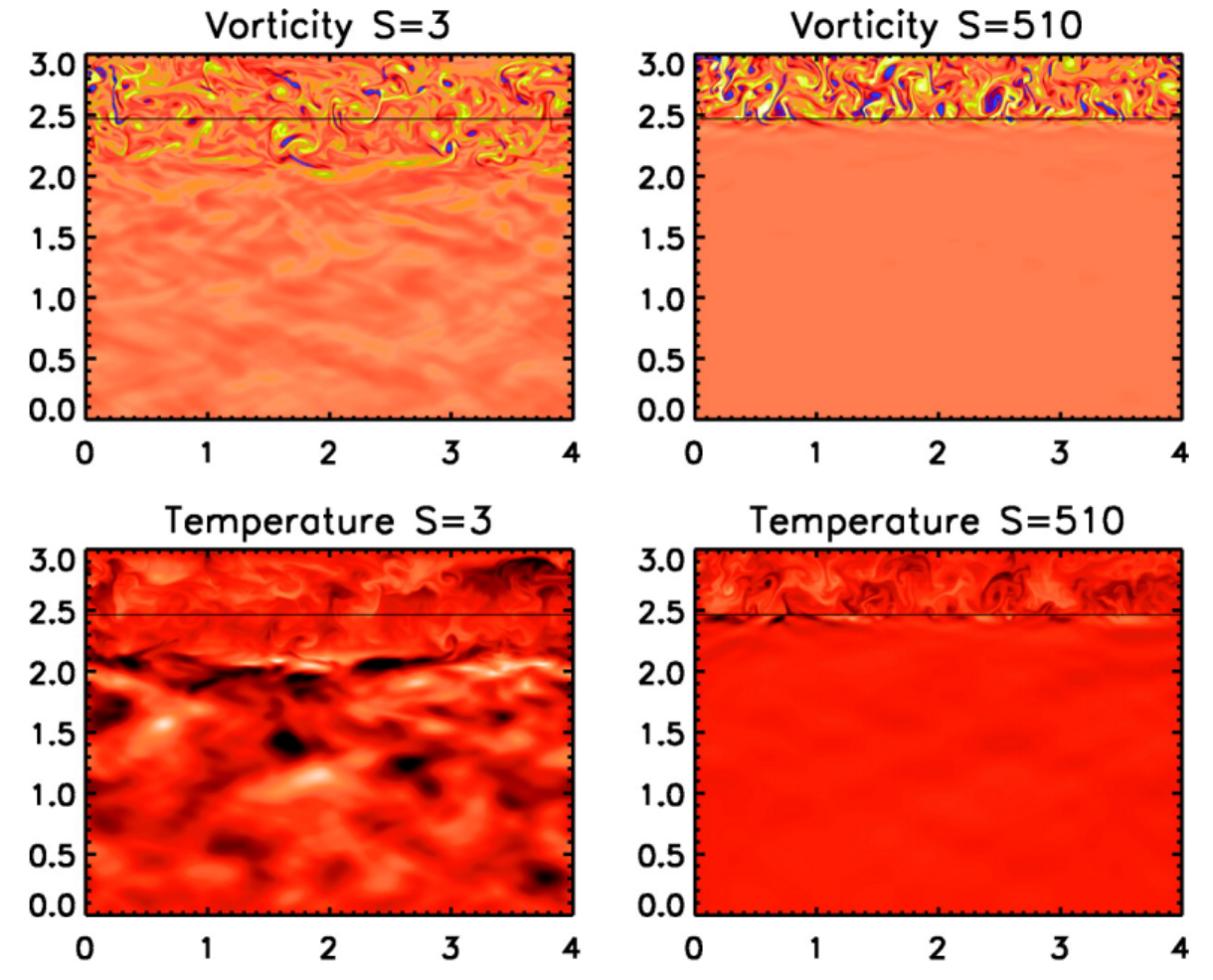}
\caption{
The vorticity (top) and temperature (bottom) are shown for simulations whose values of $\mS_{\rm struct}$ are small (left) and large (right).
The black line marks the Schwarzschild boundary of the simulations.
Note that increasing $\mS$ decreases the depth to which dynamics overshoot beyond the Schwarzschild boundary.
Note also that as stiffness increases, the internal gravity wave amplitude in the stable region decreases with respect to the convection amplitude. 
Figure~2 of \citet{rogers_glatzmaier_2005}; © AAS. Reproduced with permission.
\label{fig:overshoot_dynamics}}
\end{figure}

\emph{Convective overshoot} is a process which occurs on the scale of an individual convective flow when the flow traverses the convective boundary.
We define the convective boundary as the location where the entropy gradient becomes positive.
In the bulk convection zone, buoyancy forces act in the expected sense (low entropy blobs accelerate upwards and high entropy blobs accelerate downwards).
Beyond the convective boundary, buoyancy forces act in the opposite sense and motions become wave-like (low entropy blobs are accelerated downwards, and vice versa).
A ``hot'' upflow in the convection zone therefore accelerates downward after passing the convective boundary.
This wave-like restoring motion of a convective parcel beyond the convective boundary is convective overshoot.
Convective overshoot is visualized in Figure~\ref{fig:overshoot_dynamics}.

Convective overshoot occurs in all simulations which include a convection zone bordered by a radiative zone.
First seen by \citet{hurlburt_etal_1986}, many studies have observed overshooting and have generally sought to understand how it depends on $\mathrm{Pe}$ and $\mS_{\rm struct}$  (e.g.,~\citep{hurlburt_etal_1994,kiraga_etal_1995,bazan_arnett_1998,brummell_etal_2002,rogers_glatzmaier_2005,dietrich_wicht_2018,cai2020a,cai2020}).
Others sought to understand how the imposed convective flux determined the depth of overshoot \citep{singh_etal_1998,tian_etal_2009,kapyla2019}.

Recently, an energetics-based model of convective overshoot has emerged.
This model is laid out in \citet{korre_etal_2019}, eqns.~30-35, and is also described in \citet{lecoanet_etal_2016}.
They argue that a convective blob passing the convective boundary will overshoot adiabatically until the parcel's kinetic energy is converted into potential energy.
This argument was presented in the context of a simplified, Boussinesq model; here we briefly recreate it in the context of the fully compressible Eqns.~\ref{eqn:continuity}-\ref{eqn:energy}.
The conversion of kinetic energy into gravitational potential energy occurs through the action of buoyant work,
\begin{equation}
    \frac{1}{2}\rho_{\rm CB} u_{\rm conv}^2 = \int_{r_{\rm CB}}^{r_{\rm CB} + \delta_{\rm ov}} F_{\rm buoy}\,dr \sim \delta_{\rm ov} \angles{F_{\rm buoy}},
    \label{eqn:ke_is_pe}
\end{equation}
where $\rho_{\rm CB}$ is the density near the convective boundary, $u_{\rm conv}$ is the bulk convective velocity, $r_{\rm CB}$ is the radial location of the convective boundary, $\delta_{\rm ov}$ is the overshoot distance, and $F_{\rm buoy}$ is the radial component of the buoyancy force.
In the limit of low-Mach number flows applicable to core convection or deep envelope convection on the main sequence \citep{jermyn_etal_2022_atlas}, and for an ideal gas \citep{brown_etal_2012}, the buoyant force in this limit becomes
\begin{equation}
    \vec{F}_{\rm buoy} = -\rho \vec{g} \frac{s'}{c_p},
\end{equation}
where $s'$ is the specific entropy fluctuation associated with a convective blob and $c_p$ is the specific heat at constant pressure.
The buoyant force near the convective boundary for a parcel traveling adiabatically is approximately
\begin{equation}
    \angles{F_{\rm buoy}} \sim \int \rho \frac{g}{c_p}\frac{d s}{dr} dr = \int \rho N^2 dr.
\end{equation}
Assuming that the density is roughly the background density and does not vary sharply near the convective boundary provides
\begin{equation}
    \frac{1}{2}\rho_{\rm CB} u_{\rm cz}^2 \sim \delta_{\rm ov} \rho_{\rm CB} \int N^2 dr.
    \label{eqn:pre_integ}
\end{equation}
Dividing by a characteristic length scale $L$, $\rho_{\rm CB}$, and $\Delta b = \int N^2 dr$ provides
\begin{equation}
    \frac{\delta_{\rm ov}}{L} \sim \Ri^{-1},
\end{equation}
i.e., the overshoot depth is inversely proportional to the Richardson number or similarly the dynamical stiffness $\mS_{\rm dyn}$.
\citet{korre_etal_2019} take this argument one step further for more direct comparison with simplified models of overshoot in convective simulations.
They derive how they expect the overshoot distance to scale with $\mS_{\rm dyn}$.
For a stratification near the convective boundary like $N^2 = N_0^2 [(r-r_{\rm cb})/L]^{\alpha}$ (where $r_{\rm cb}$ is the radial coordinate of the convective boundary), they evaluate $\Delta b$ and rearrange Eqn.~\ref{eqn:pre_integ} to find
\begin{equation}
    \Delta b = \frac{N_0^2}{1 + \alpha} \frac{\delta_{\rm ov}^{1+\alpha}}{L^\alpha}
    \,\,\Rightarrow\,\,
    \frac{1 + \alpha}{2}\frac{u_{\rm cz}^2}{L^2 N_0^2} \sim \left(\frac{\delta_{\rm ov}}{L}\right)^{2 + \alpha}
    \,\,\Rightarrow\,\,
    \left(\frac{\delta_{\rm ov}}{L}\right) \sim \mS_{\rm dyn}^{-1/(2 + \alpha)}.
\end{equation}
So e.g., for a simulation where $N^2$ is a constant above the convective boundary, $(\delta_{\rm ov}/L) \sim \mS_{\rm dyn}^{-1/2}$, and for a simulation where it increases linearly, it should scale as $(\delta_{\rm ov}/L) \sim \mS_{\rm dyn}^{-1/3}$, which is the case examined by \citet{lecoanet_etal_2016}.
A stratification which is well approximated by $N^2 \sim r^2$ in a small region outside of the convection zone would therefore reproduce the $\mS^{1/4}$ observed by early simulations \citep{hurlburt_etal_1994,brummell_etal_2002}.

We note briefly that one fairly robust result from the literature of overshooting convection is that overshoot depths are almost universally seen to decrease as $\mS_{\rm struct}$ (and $\mS_{\rm dyn}$) increases.
One exception is the recent result of \citet{cai2020} who observed that convective exit velocities increased in the very-high $\mS_{\rm struct}$ regime, which in turn produced increasing overshoot depths; the process which would drive these increased velocities is not clear.


\subsection{Convective Overshoot as Turbulent Diffusion}
\label{subsec:3d_turb_diff}

Overshooting has been incorporated into 1D stellar evolution models by parameterizing convective mixing using a diffusivity profile (see Section~\ref{subsec:mixing_profs}).
An exponential diffusive profile was observed in the early simulations of \citet{Freytag1996A&A...313..497F}; this profile was adopted by \citet{Herwig2000A&A...360..952H}, and this has been the standard choice in the field ever since.
More recently, \citet{jones_etal_2017} found that an exponential turbulent diffusivity described the turbulent diffusion measured in their simulations well.
\citet{herwig_etal_2006} have noted that convective velocities start to fall off before reaching the CZ boundary, which complicates the implementation of this exponential diffusivity.
Separately, \citet{lecoanet_etal_2016} see a fast decrease of turbulent diffusivity outside of their convection zone, but they argue that it is better parameterized as \emph{step} overshooting (Section~\ref{subsec:Overshoot}).

We note that there are two separate questions which must be answered to robustly describe convective overshoot as a turbulent diffusivity.
First, how do the convective \emph{velocities} decrease beyond the convective boundary?
\citet{korre_etal_2019} find that the kinetic energy is well-defined by a Gaussian beyond the convective boundary (e.g., their fig 4), while \citet{Pratt2017A&A...604A.125P} use extreme value statistics to characterize the \emph{maximum} depth that convective plumes overshoot to at any point in time, and find their results best-described by a Gumbel distribution ($\mathrm{exp}(-\mathrm{exp}(-x))$).
Once the velocity profile beyond the convective boundary is understood, we must then ask how the velocity profile relates to the \emph{mixing} produced by overshooting convection.

\subsection{Entrainment}
\label{subsec:3d_entrainment}
\emph{Entrainment} is the process by which convection ``scrapes'' material from an adjacent stable layer into the convective region and then mixes that material.
Entrainment is caused by multiple processes (e.g., splashing from convective overshoot or shear instabilities driven by horizontal convective flow \citep{woodward_etal_2015}); for simplicity, we consider entrainment to be any process accompanied by a measurable mixing of the mean radial entropy or chemical profile at the convective boundary.
Energetically, entrainment occurs when convection exerts buoyancy work \emph{on} the adjacent stable fluid.
This work raises the potential energy of a portion of the adjacent fluid enough to dislodge it and drag it into the convecting region.

The earliest entrainment studies examined a stable, linear composition gradient which was destabilized by heating from below, resulting in the emergence of a convection zone which grows by entrainment.
This simple setup has been studied both in the lab and numerically for the past 60 years \cite{turner_1968,deardorff_etal_1969,kato_phillips_1969,linden_1975,fernando_1987,molemaker_dijkstra_1997,leppinen_2003,fuentes_cumming_2020}.
These studies proposed and observed an $E \propto \Ri^{-1}$ relationship, where $E = u_e/u_t$ is the entrainment efficiency, with $u_e$ the entraiment velocity (the rate at which the convective boundary advances) and $u_t$ is the turbulent convective velocity \citep{kato_phillips_1969}.
These studies measured the height of the convective boundary vs.~time $h(t)$, and found generally $h(t) \propto t^{1/2}$ (e.g.,~\citep{turner_1968,fernando_1987,fuentes_cumming_2020}) or $h(t) \propto t^{1/3}$ \citep{kato_phillips_1969}.
Entrainment has also been observed and studied in other simulations of Boussinesq convection bounded by a stable region; these studies further established the dependence of the entrainment rate on $\Ri$ or $\mS$ \cite{couston_etal_2017,toppaladoddi_wettlaufer_2018}.
 
\begin{figure}[t]
\centering
\includegraphics[width=\linewidth]{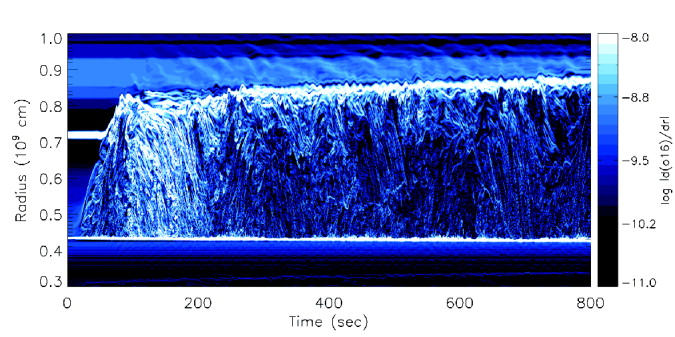}
\caption{
Data are from a hydrodynamical simulation of an oxygen-burning shell; radial coordinate is on the y-axis and time coordinate is on the x-axis.
Color shows the radial gradient of the oxygen concentration profile at each time; the thick bright lines denote the top and bottom boundaries of the convective region.
Turbulent convection occurs at times $t \gtrsim 50$, and entrainment causes measurable movement of the convective boundary. Figure~4 of \citet{meakin_arnett_2007}; © AAS. Reproduced with permission.
\label{fig:entrainment_dynamics}}
\end{figure}

\begin{figure}[t]
\centering
\includegraphics[width=0.7\linewidth]{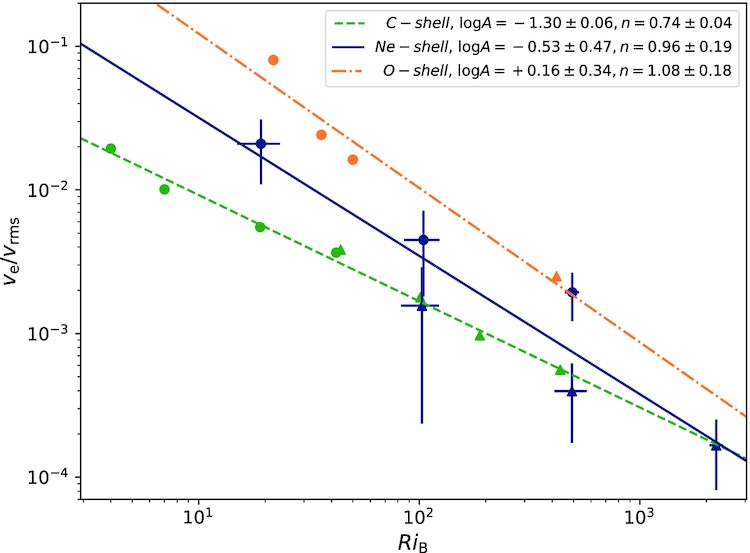}
\caption{
Entrainment rate is plotted against the bulk Richardson number for simulations of neon burning shells \citep[][]{rizzuti_etal_2022}, oxygen burning shells \citep[][]{meakin_arnett_2007}, and carbon burning shells \citep{cristini_etal_2019}.
Entrainment laws $A Ri_{\rm B}^{-n}$ with $A \sim (0.05, 1)$ and $n \sim 1$ are reported. Figure~5 of \citet{rizzuti_etal_2022}; © Oxford University Press. Reproduced with permission.
\label{fig:entrainment_scaling}}
\end{figure}

More recently, \citet{meakin_arnett_2007} introduced the concept of turbulent entrainment into studies of stellar astrophysics.
They perform 3D hydrodynamic simulations of stellar convection using stellar structure models as initial conditions and find significant turbulent entrainment and advancement of the convective boundary (see Figure~\ref{fig:entrainment_dynamics}).
They find that the entrainment efficiency follows a power law scaling of $E = A \Ri^{-n}$, where $A \sim 10^{-1} - 10^{-2}$ and $n \approx 1$, well in line with previous geophysical studies.
These results have been corroborated by many hydrodynamical studies over the past decade (e.g.,~\citep{arnett_etal_2009,mocak_etal_2009,gilet_etal_2013,jones_etal_2017,cristini_etal_2019,rizzuti_etal_2022}), typically finding power laws with $n \approx 1$ and $A \in [10^{-2},1]$ (see Figure~\ref{fig:entrainment_scaling}).
These results have inspired \citet{staritsin_2013} and \citet{scott_etal_2021} to include power-law implementations of turbulent entrainment into stellar models, but they find that the entrainment law calibrated to simulations leads to the entire star being engulfed by the convection zone on evolutionary timescales.
They do find decent agreement with other forms of boundary mixing using $A \sim 10^{-4}$, but to date no dynamical simulations have revealed a value of $A$ this small.

We interpret the state of the astrophysical entrainment literature as follows.
Stellar models \emph{underestimate} the size of convection zones consistently.
As a result, when stellar models are used as initial conditions for 3D hydrodynamical simulations, significant turbulent entrainment is observed as the convecting regions expand to an equilibrium size.
Unfortunately most simulations are not long enough to observe the equilibrium sizes of convecting regions, so the saturation size of convective zones is uncertain.
\citet{anders_etal_2022b} studied a simulation under the Boussinesq approximation in which the Ledoux and Schwarzschild criteria initially disagree regarding the location of the convective boundary.
Convection entrains material at the Ledoux boundary until the two criteria agree, after which point entrainment stops.
Unfortunately, we are unaware of any studies which both employ the fully compressible equations and allow the size of the convection zone to fully saturate through entrainment, so these findings should be confirmed in more complex setups.

One may ask if entrainment should be included in standard stellar evolution models, just like exponential and step overshoot prescriptions.
We believe that a precise implementation of entrainment is not necessary during the main sequence or other phases of evolution where the evolutionary timescale is very long compared to the entrainment rate \citep{anders_etal_2022b}.
However, proper entrainment implementations will improve stellar evolution calculations of short-lived phases of evolution where the size of convection zones are changing rapidly and where time-dependent convection implementations are necessary \citep{jermyn_etal_2022_mesa6}.



\subsection{Convective Penetration}
\label{subsec:3d_penetration}
The boundary of a well-mixed convective region can advance by entrainment significantly beyond the Schwarzschild boundary.
When this occurs, we refer to the process as \emph{convective penetration}, characterized by a nearly adiabatic and chemically homogeneous region which is part of the convection zone but which is characterized by $\gradrad < \gradad$.

Convective penetration was hypothesized by Roxburgh and Zahn \cite{roxburgh_1978,roxburgh_1989,roxburgh_1992,Zahn1991A&A...252..179Z}, but has been elusive in simulations and experiments.
The hallmark of penetrative convection is mixing of the entropy gradient beyond the Schwarzschild boundary.
Entropy mixing beyond the \emph{initial} convective boundary has often been reported \citep{hurlburt_etal_1994,saikia_etal_2000,brummell_etal_2002,rogers_glatzmaier_2005,rogers_etal_2006,kitiashvili_etal_2016,baraffe_etal_2021}, but it is often unclear if the reported process is convective penetration or if it is movement of the Schwarzschild boundary by entrainment.
Another hallmark of penetrative convection is substantial negative convective flux (and excess radiative flux) beyond the schwarzschild boundary; this is frequently observed \cite{hurlburt_etal_1986,singh_etal_1995,browning_etal_2004,brun_etal_2017,Pratt2017A&A...604A.125P}, but also often seen in studies of non-penetrative overshooting convection.

Unfortunately, studies aimed at understanding convective penetration have found inconsistent or contradictory results regarding how penetration depends on e.g., $\mS$ or $\Ri$.
Early studies \citep{hurlburt_etal_1994,singh_etal_1995} suggested that penetration and $\mS_{\rm struct}$ were linked at low values of $\mS_{\rm struct}$ (a regime that produces a dynamical $\mS_{\rm dyn}$ which is not relevant for core convection, see \citet{couston_etal_2017}), but later studies \citep{brummell_etal_2002,rogers_glatzmaier_2005} found no link between $\mS_{\rm struct}$ and convective penetration.
However, many simulations have found that penetration lengths can depend on the magnitude of energy fluxes \cite{singh_etal_1998,kapyla_etal_2007,tian_etal_2009,hotta2017,kapyla2019}.

\begin{figure}[t]
\includegraphics[width=\linewidth]{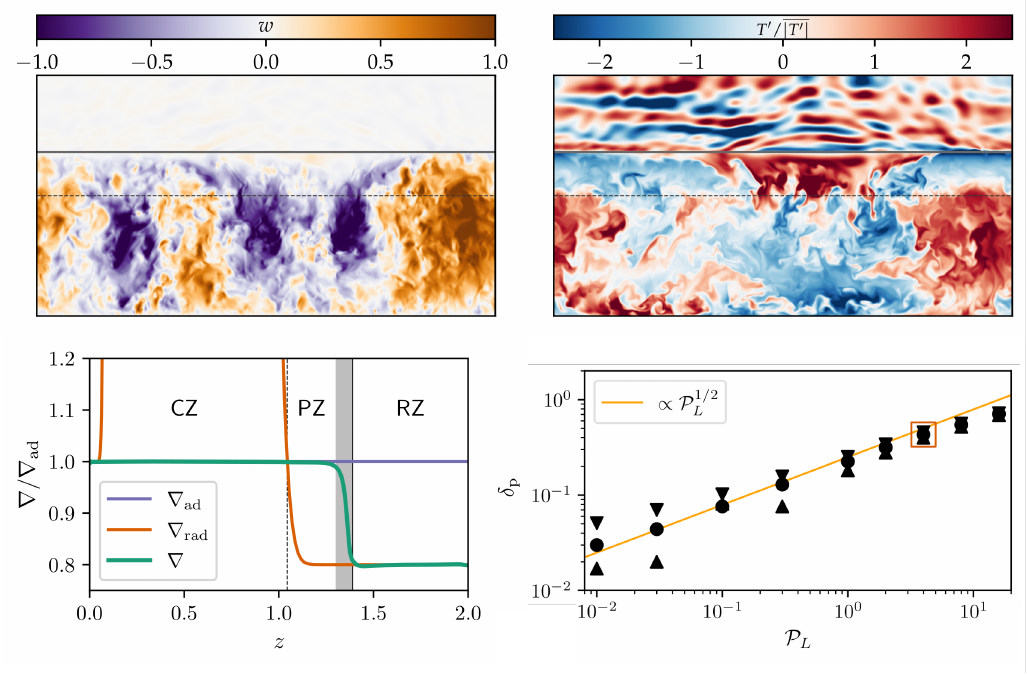}
\caption{
A snapshot in time of a 3D hydrodynamic simulation is displayed in the upper two panels with the velocity (left) and scaled temperature anomaly (right) shown.
The Schwarzschild boundary is shown as a dashed line, and the height where $\grad$ departs from $\gradad$ is plotted as a thick horizontal line.
Note in the upper right panel that hot CZ upwellings turn cold in the penetration zone between the dashed and solid lines.
(Bottom left) The time- and horizontally-averaged $\grad$ profiles from a simulation like the one in the top panel; a distinct convection zone (CZ), nearly adiabatic penetration zone (PZ), overshoot zone (grey shaded region) and radiative zone (RZ) are seen.
(Bottom right) The extent of the PZ is plotted against the penetration parameter, with the expected scaling overplotted in orange; $\mathcal{P}$ is found to be the most important parameter for determining the size of the PZ. 
From Figure~1 (upper two panels), Figure~2. (bottom left panel) and part of Figure~7 (bottom right panel) from \citet{anders_etal_2022a}; these figures were originally published online under a CC BY license.
\label{fig:hydro_penetration} }
\end{figure}

Robust evidence of convective penetration in numerical simulations was observed by \citet{anders_etal_2022a} in 3D Cartesian and \citet{baraffe_etal_2023} in 2D spherical simulations.
The dynamics of penetrative convection are shown in the top two panels of Figure~\ref{fig:hydro_penetration}.
The thermal structure of a convective region with a penetration region is shown in the bottom left panel of Figure~\ref{fig:hydro_penetration}.
The extent of the penetration region scales strongly with the penetration parameter $\mathcal{P}$, as shown in the bottom right panel of Figure~\ref{fig:hydro_penetration}.

Convective penetration occurs in the stationary state, so Eqn.~\ref{eqn:therm_integral_template} becomes
\begin{align}
    -\angles{\frac{1}{T}\left(\grad\cdot(k\grad T) + \rho \epsilon\right)} &= \angles{\frac{1}{T}\Phi}. \label{eqn:thermal_diss}
\end{align}
This can be rearranged into the integral constraint of \citet{roxburgh_1989,roxburgh_1978,roxburgh_1992},
\begin{equation}
    \int_V -(F_{\rm tot} - F_{\rm rad})\frac{1}{T_0^2}\frac{d T_0}{dr} dV 
    = \int_V \frac{\Phi}{T_0} dV,
    \label{eqn:roxburgh}
\end{equation}
where $F_{\rm tot}$ is the total flux, $F_{\rm rad}$ is the radiative flux, and $T_0$ is the temperature stratification.
We follow \citet{anders_etal_2022b} and break up constraint integrals into a CZ (convective zone) and PZ (penetrative zone) portion.
Noting that $F_{\rm conv} = F_{\rm tot} - F_{\rm rad}$, and that $d T_0 / dr < 0$, we get
\begin{equation}
    \int_{\rm CZ} F_{\rm conv} \bigg|\frac{1}{T_0^2}\frac{d T_0}{dr}\bigg| dV = 
    \int_{\rm CZ} \frac{\Phi}{T_0} dV
    + \left[
    -\int_{\rm PZ} F_{\rm conv} \bigg|\frac{1}{T_0^2}\frac{d T_0}{dr}\bigg| dV
    + \int_{\rm PZ} \frac{\Phi}{T_0} dV \right].
    \label{eqn:integ_constraint}
\end{equation}
The left-hand side (LHS) of Eqn.~\ref{eqn:integ_constraint} is a buoyant ``engine'' which quantifies the buoyant work done by the convection in the bulk CZ.
In a energetically stationary state, this positive work must be balanced out by the terms on the right-hand side (RHS) of the equation.
We note that in an \emph{adiabatically mixed PZ}, where $\grad \approx \gradad$ but $\gradad > \gradrad$, radiation carries too much flux $F_{\rm rad} > F_{\rm tot}$, so $F_{\rm conv} < 0$ is required for equilibrium.
Therefore, all terms on the RHS of Eqn.~\ref{eqn:integ_constraint} are positive and contribute to consumption of the LHS work.
We therefore see that either dissipation is highly efficient in the convection zone, or a penetrative region characterized by negative buoyant work is required to achieve energy conservation.
Zahn \cite{Zahn1991A&A...252..179Z} noted that the size of a penetrative region is controlled by how drastically $\gradrad$ departs from $\gradad$ in the PZ.
This intuition appears mathematically here: a rapid departure where $\gradrad \ll \gradad$ (small $\mathcal{P}$) leads to a large negative $F_{\rm conv}$, so only a small PZ is required for balance in Eqn.~\ref{eqn:integ_constraint}.
A very gradual departure where $\gradrad \sim \gradad$ (large $\mathcal{P}$) leads to $F_{\rm conv} \sim 0$ in the PZ, and so its size is set by the turbulent dissipative properties of the convection.

We note that it is unintuitive for dissipation ($\Phi$) to play an important role in astrophysical convection, where viscosities are very small.
In turbulent flows, the so-called ``zeroth law of turbulence'' states that energy which is injected into a turbulent cascade at large scales must eventually be dissipated at small scales.
Therefore the rate of turbulent dissipation is not determined by the magnitude of viscosity but rather by the rate of energy transfer from the largest eddies into the cascade, which scales something like $U^3/\ell$ (Ref.~\citep{pope_turbflows}, Section 6.1.1).
We also note that astrophysical convection occurs in a magnetized plasma, where additional dissipation processes (e.g., Ohmic dissipation) complicate this picture, but a full discussion of this work is beyond the scope of this review.
We simply note that dissipation is expected to be substantial, and this has been shown in direct numerical simulations (e.g.,~\citep{singh_etal_1995,viallet_etal_2013,currie_browning_2017}), although a satisfying model for the magnitude of viscous dissipation in astrophysical convection has not yet been created.
This is an idea that appears both in the literature of convective penetration, and in the other fields like gravity wave mixing \citep{kupka_etal_2022}, and should be examined in more detail.


\subsection{Rotational Constraint \& Magnetic Pumping}
\label{subsec:3d_rotation_magnets}
In this review, we focused on results from hydrodynamical (not \emph{magneto}hydrodynamical), nonrotating fluid simulations.
All stars rotate and this rotation often strongly influences convective dynamics \citep{jermyn_etal_2022_atlas}.
It is generally believed that rotation should \emph{decrease} the extent of CBM (although it may create e.g., meridional circulations which themselves separately increase mixing).
There is evidence that rotation increases the dissipation in convective flows \citep{julien_etal_1996,julien_etal_2012,aurnou_etal_2020}, which would decrease the extent of a penetration zone; analytic work by \citet{Augustson2019ApJ...874...83A} also predicts that rotation should decrease the extent of penetration zones.
\citet{brummell_etal_2002} find that rotation decreases overshoot, while \citet{dietrich_wicht_2018} find that only $\mS$ affects overshoot and not rotation.
\citet{browning_etal_2004} studied 3D rotating core convection and found prolate penetration zones aligned with the rotation axis, which is consistent with the local box simulations of \citet{pal_etal_2008} who found less penetration at the equator than at the poles.
The effects of magnetism are even less studied, but convection can pump magnetic fields out of the convection zone and into CBM regions; this process was discovered by \citet{drobyshevski_yuferev_1974} and has been observed in simulations \citep{tobias_etal_2001,ziegler_rudiger_2003}; magnetic pumping has been suggested as a mechanism for solar active region formation \citep{fisher_etal_1991} and has been used to study the structure of the Sun's magnetic field below the convection zone \citep{vanballegooijen1982}.
The manner in which rotation and magnetism affect CBM remains unclear, and future studies should explore the importance of these effects on each of the processes discussed in this review.

\section{Empirical calibrations}
\label{sec:observations}

\subsection{Stellar clusters}
\label{subsec:clusters}

Early observational inferences on CBM dating back to 1971 come from the studies of the old open cluster M67. Both \citet{Racine1971ApJ...168..393R} and \citet{Torres-Peimbert1971BOTT....6....3T} reported that the observed gap above the main-sequence turnoff, could not be reproduced by isochrones computed with standard models \cite{Torres-Peimbert1971BOTT....6....3T}. 
A hook is seen in the isochrone at the main-sequence turnoff and the rapid evolution caused by hydrogen exhaustion results in a gap in the number of stars observed above the turnoff. 
By including CBM in the models, \citet{Prather1974ApJ...193..109P} demonstrated that this hook and hence gap can persist to much greater ages and thereby explain the observed gap in M67. Similar discrepancies between observations and standard model predictions were found around the same time for the open clusters NGC~752 \cite{Bell1972MNRAS.157..147B} and NGC~2420 \cite{McClure1978PASP...90..170M}. For the latter cluster, an update in the adopted opacity tables remained insufficient to explain the gap and the inclusion of CBM in the models was required \cite{Demarque1994ApJ...426..165D}. In comparison, \citet{Maeder1981A&A....93..136M} considered a sample of 34 open clusters, finding that the inclusion of CBM is required to explain the extension of the core-hydrogen burning phase beyond the theoretical sequence predicted by standard models.

Several additional open and globular clusters have been studied in detail to investigate whether CBM is required to explain their morphology and distribution of stars in the color-magnitude diagram (CMD) (e.g. NGC~3680 \cite{Kozhurina-Platais1997AJ....113.1045K}, IC~4651 \cite{Andersen1990ApJ...363L..33A}, NGC~2164 \cite{Vallenari1991A&AS...87..517V}, NGC~1831 \cite{Vallenari1992AJ....104.1100V}, NGC~1866 \cite{Chiosi1989A&A...219..167C}, NGC~6134 \cite{Bruntt1999A&AS..140..135B}, NGC~2173 \cite{Woo2003AJ....125..754W}, SL~556 \cite{Woo2003AJ....125..754W}, NGC~2155 \cite{Woo2003AJ....125..754W}, NGC~1783 \cite{Mucciarelli2007AJ....134.1813M}, NGC~419 \cite{Girardi2009MNRAS.394L..74G}). Meaningful inferences on the CBM from these types of studies requires non-cluster and binary members to be properly identified \cite{Andersen1990ApJ...363L..33A,Nordstrom1991Msngr..63...34N}. 
Isochrones including CBM improve model agreement with observations for these clusters. The inclusion of improved opacity tables in the models generally tend to decrease the amount of required CBM, and in some cases may be sufficient to explain the observations without requiring any additional CBM \cite{Dinescu1995AJ....109.2090D}. 

\citet{Demarque2004ApJS..155..667D} were some of the first to consider a mass dependent CBM in the calculation of model isochrones. The \texttt{$Y^2$} isochrones included a gradual increase in the CBM parameter up to a critical mass above which a constant value was assumed, finding good fits to the observed CMDs of the seven considered open clusters including M67. 
However, the need for CBM is not unambiguous. \citet{Michaud2004ApJ...606..452M} argued that no CBM is required to reproduce the observed CMD of either M67 or NGC~188 if microscopic diffusion is included in the models. A similar conclusion for M67 was later found by \citet{Viani2017EPJWC.16005005V}.

Similar mass dependent CBM to the one adopted by \citet{Demarque2004ApJS..155..667D} has since been included in other isochrones such as the \texttt{PARSEC} isochrones \cite{Bressan2012MNRAS.427..127B}, while others like the \texttt{MIST} isochrones \cite{Dotter2016ApJS..222....8D,Choi2016ApJ...823..102C} adopt a single value for the CBM parameter. Recently, \citet{Johnston2019MNRAS.482.1231J} introduced the concept of an isochrone cloud, which shows what an isochrone would look like if the internal mixing were allowed to vary on a star-by-star basis without assuming, e.g., a mass-dependent CBM. In this case the isochrone is no longer a thin line but fans out for masses with convective cores. Same age models with higher mixing will be less evolved due to the extended main-sequence lifetime and define the blue edge of the isochrone cloud, whereas models with less mixing will be further evolved and therefore have lower temperatures corresponding to the red edge of the isochrone cloud, see Figure~3 of \citet{Johnston2019A&A...632A..74J}. The isochrone clouds were
later used to model two younger stellar clusters showing extended main-sequence turnoffs (eMSTOs)\cite{Johnston2019A&A...632A..74J}. eMSTOs are a broadening of the main-sequence of a cluster near its turn-off for $M \gtrsim 1.4\,\text{M}_\odot$, and are common in young and intermediate age clusters (e.g.,~\citep{Milone2018MNRAS.477.2640M,Goudfrooij2018ApJ...864L...3G}). 
Age spreads \cite{Milone2009A&A...497..755M}, binary interactions \cite{Yang2011ApJ...731L..37Y}, rotation \cite{Bastian2009MNRAS.398L..11B}, and variations in CBM \cite{Yang2017ApJ...836..102Y} have been suggested as possible explanations for the eMSTOs. Spectroscopic observations focusing on measuring projected rotational velocities, $v\sin i$, of stars in the eMSTO have shown in recent years that the spread appears to coincide with a spread in $v\sin i$ amongst the stars, with faster rotating stars being redder and cooler than those with lower projected rotational velocities \cite{Bastian2018MNRAS.480.3739B,Marino2018AJ....156..116M,Sun2019ApJ...876..113S,Kamann2020MNRAS.492.2177K,Kamann2023MNRAS.518.1505K}. 
These observations suggest that rotation is the dominant effect behind the eMSTO, and that the eMSTO is caused by the combined effects of gravity darkening \cite{vonZeipel1924MNRAS..84..665V,EspinosaLara2011A&A...533A..43E}, where rotation causes the equators of the stars to be cooler than the poles, and a spread in inclination angles. \citet{Lipatov2022ApJ...934..105L} recently provided a tool for accounting for these effects in the model isochrones. Note that the effects on the positions of the stars in the CMD caused by gravity darkening and spreads in inclination angles are opposite to those caused by internal mixing, where faster rotating stars are expected to have higher amounts of internal mixing. Knowing the inclination angles of the stars could help disentangling the relative importance of these different effects on the morphology of the eMSTO.



\subsection{Apsidal motion}
\label{subsec:apsidal}

Apsidal motion, the change in the position of the periastron of a binary orbit, provides direct evidence of the internal density concentration of the stars in the binary system \cite{Russell1928MNRAS..88..641R}. Measurements of Apsidal motion are based on the calculation of the apsidal constant $k_j$ ($j=2,3,4$), also known as the density or internal structure constant. From an observational standpoint, only the second order apsidal motion constant $k_2$ is usually important \cite{Claret1991A&A...244..319C}, and takes on a value of $k_2 = 0.75$ for a homogeneous density distribution \cite{Russell1928MNRAS..88..641R,Claret1991A&A...244..319C}. 
In reality, rather than deriving the individual component apsidal constants, one instead works with a weighted average value $\overline{k_2}$ of the two binary components  \cite{Claret1991A&A...244..319C}.

\begin{figure}[t]
\includegraphics[width=\linewidth]{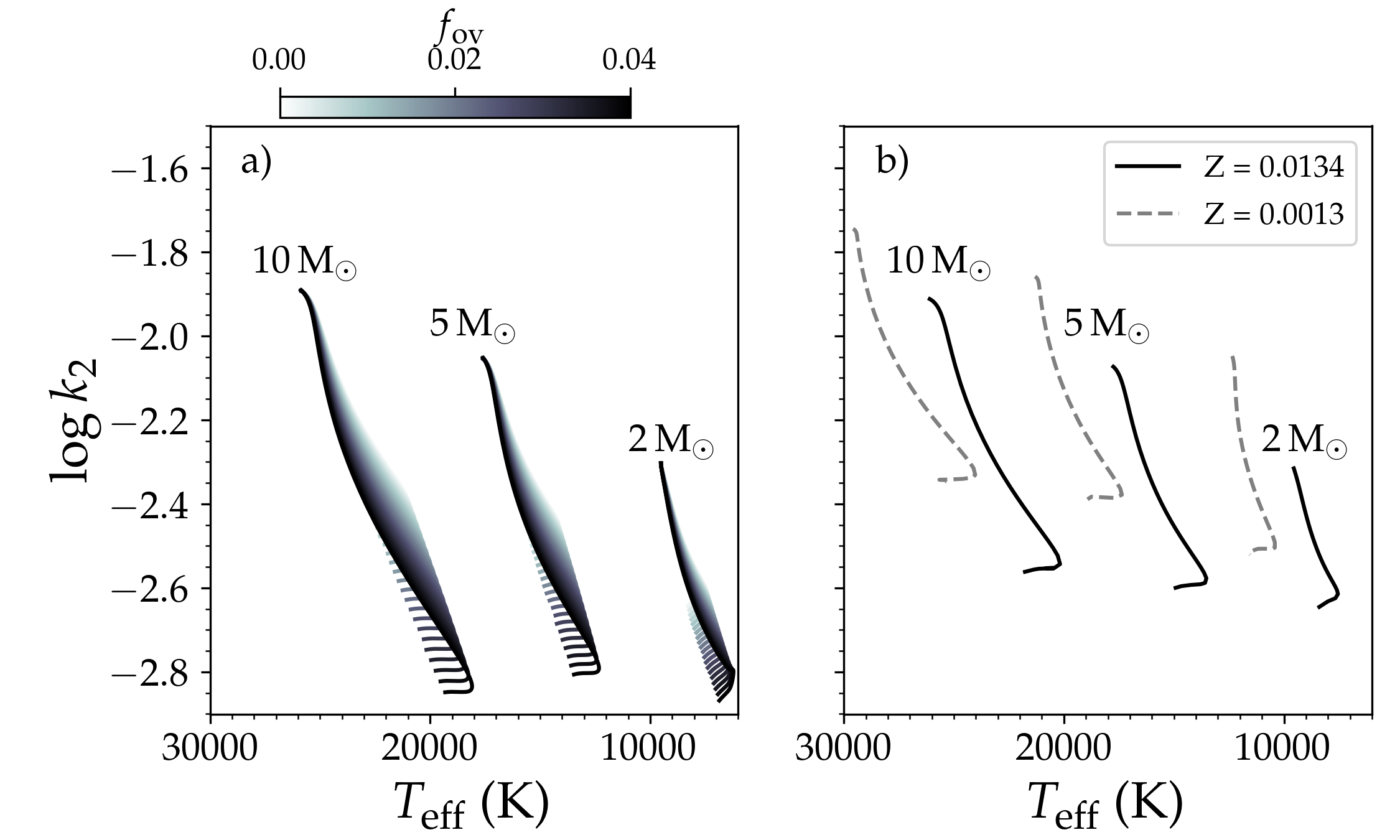}
\caption{Evolution of the apsidal constant $k_2$ for three different initial stellar masses. (\textbf{a}) Variation in $k_2$ resulting from varying the extent of the CBM region assuming exponential diffusive overshoot and a fixed initial chemical composition of  $X = 0.71$, $Y=0.276$, and $Z=0.014$. (\textbf{b}) Variation in $k_2$ resulting from changing the initial metallicity of the stars, assuming a mass dependent CBM \cite{Claret2019A&A...628A..29C}. Figure made by the authors using \texttt{MESA} models (panel \textbf{a}) and preexcisting model grids by \citet{Claret2019A&A...628A..29C} (panel \textbf{b}). \texttt{MESA} inlists and data used to generate panel (\textbf{a}) are available on Zenodo \citep{supp}.
\label{fig:logk2_CBM}}
\end{figure}

Figure~\ref{fig:logk2_CBM} a) illustrates how the size of the CBM region affects $\log k_2$ throughout the main-sequence evolution for three different initial stellar masses. The decrease in $\log k_2$ during the main-sequence evolution is caused by the fusion of hydrogen to helium, resulting in the stars becoming more centrally condensed as they evolve. Aside from CBM, changing the opacity and metallicity of the models likewise changes the predicted $k_2$ values \cite{Semeniuk1968AcA....18...33S}, see panel b of Figure~\ref{fig:logk2_CBM}. 
Stellar rotation also impacts the derived $k_2$ values by making the stars more centrally condensed \cite{Stothers1974ApJ...194..651S}. Finally, both CBM and stellar winds lead to more centrally condensed models but impact the stellar luminosities differently by making the models more (less) luminous when CBM (mass loss) is included \cite{Claret1991A&A...244..319C}.

The binary system Spica ($\alpha$~Virginis) is one of the first systems where the measured apsidal motion constants implied a need for CBM to reconcile models with observations  \cite{Mathis1973ApJ...180..517M,Odell1974ApJ...192..417O}. Initial studies of this system showed that the observed luminosities and effective temperatures could be matched to the models by varying the initial mass and helium content, but the predicted apsidal constant was a factor of two too high compared to the observed value. Including CBM allowed $L$, $T_\text{eff}$, and $k_2$ to simultaneously be reconciled. However, the use of different opacity tables could potentially reconcile these quantities without CBM \cite{Stothers1974ApJ...194..651S}. Further constraints could potentially be obtained by studying the primary star, which is a known $\beta$~Cep pulsator. While the oscillations of this star have previously been studied using both photometry and spectroscopy \cite{Shobbrook1969MNRAS.145..131S,Shobbrook1972MNRAS.156..165S,Smith1985ApJ...297..206S,Smith1985ApJ...297..224S,Harrington2009ApJ...704..813H,Tkachenko2016MNRAS.458.1964T}, no detailed asteroseismic modeling has so far been achieved. Given that the primary $\beta$~Cep star has been selected as a priority 1A target for the future asteroseismic CubeSpec space mission \cite{Bowman2022A&A...658A..96B} this might change in the future.

The need for CBM to reconcile the observed apsidal constants with theoretical values is not unambiguous. \citet{Claret1993A&A...277..487C} studied 14 eclipsing binaries in the mass range of 1.5-23\,M$_\odot$ and showed that a rotation correction to $k_2$ could reconcile the observed and theoretical values. For the binary system PV Cas, however, the inclusion of rotation was insufficient to reconcile the observed and theoretical $k_2$ values \cite{Claret2008A&A...490.1103C}. 
Several studies which employ a single value of the CBM parameter for all masses find good agreements between modeled and observed apsidal motion constants within the observational errors (e.g.,~\cite{Wolf2008MNRAS.388.1836W,North2010A&A...520A..74N,Bulut2013NewA...21...22B,Zasche2013A&A...558A..51Z,Lacy2015AJ....149...34L,Bakis2015NewA...40...14B,Hong2019AJ....158..185H}), while others find the theoretical values to be either larger (e.g.,~\cite{Andersen1985A&A...151..329A,Gimenez1987MNRAS.224..543G,Bakis2008MNRAS.384.1657B,North2010A&A...520A..74N,Hong2019AJ....158..185H}) or smaller (e.g.,~\cite{Degirmenci2003A&A...409..959D,Bakis2010PASJ...62.1291B,Bulut2013NewA...21...22B}) than observations. 
More recent studies rely on model grids where a mass dependent CBM was assumed (see Section~\ref{Sec:mass_dep_CBM}), and find good agreement between theory and observations \cite{Baroch2022A&A...665A..13B}. 
A few studies have tried to optimize the CBM parameters of individual binary components based on the apsidal constant. 
One such study of 27 double lined eclipsing binaries found good agreement with a predetermined mass dependent overshooting\footnote{Similar to Equation 2 of \citet{Claret2018ApJ...859..100C}, which is discussed in Section~\ref{Sec:mass_dep_CBM}.} 
The only outlier was the moderately evolved, high mass ($M_1 \approx 14$\,M$_\odot$, $M_2 \approx 11$\,M$_\odot$) system V453~Cyg, where more CBM was favored. This is not the only example of systems requiring higher CBM parameters. The study of the apsidal motion of the high mass binary system V380~Cyg indicated the need for a high overshooting parameter of $\alpha_\text{ov} \approx 0.6 \pm 0.1$  for the primary component \cite{Guinan2000ApJ...544..409G}, though the errors on the estimate was later suggested to be larger \cite{Claret2003A&A...399.1115C}. Finally, a separate study of two massive, eccentric binary systems in the open cluster NGC~6231 required enhanced internal mixing either from CBM or turbulent diffusion to reconcile the theoretical apsidal constants with the observed values \cite{Rosu2022A&A...664A..98R}.

\subsection{Mass discrepancy}
\label{subsec:mass_discrepancy}

\begin{figure}[t]
\includegraphics[width=\linewidth]{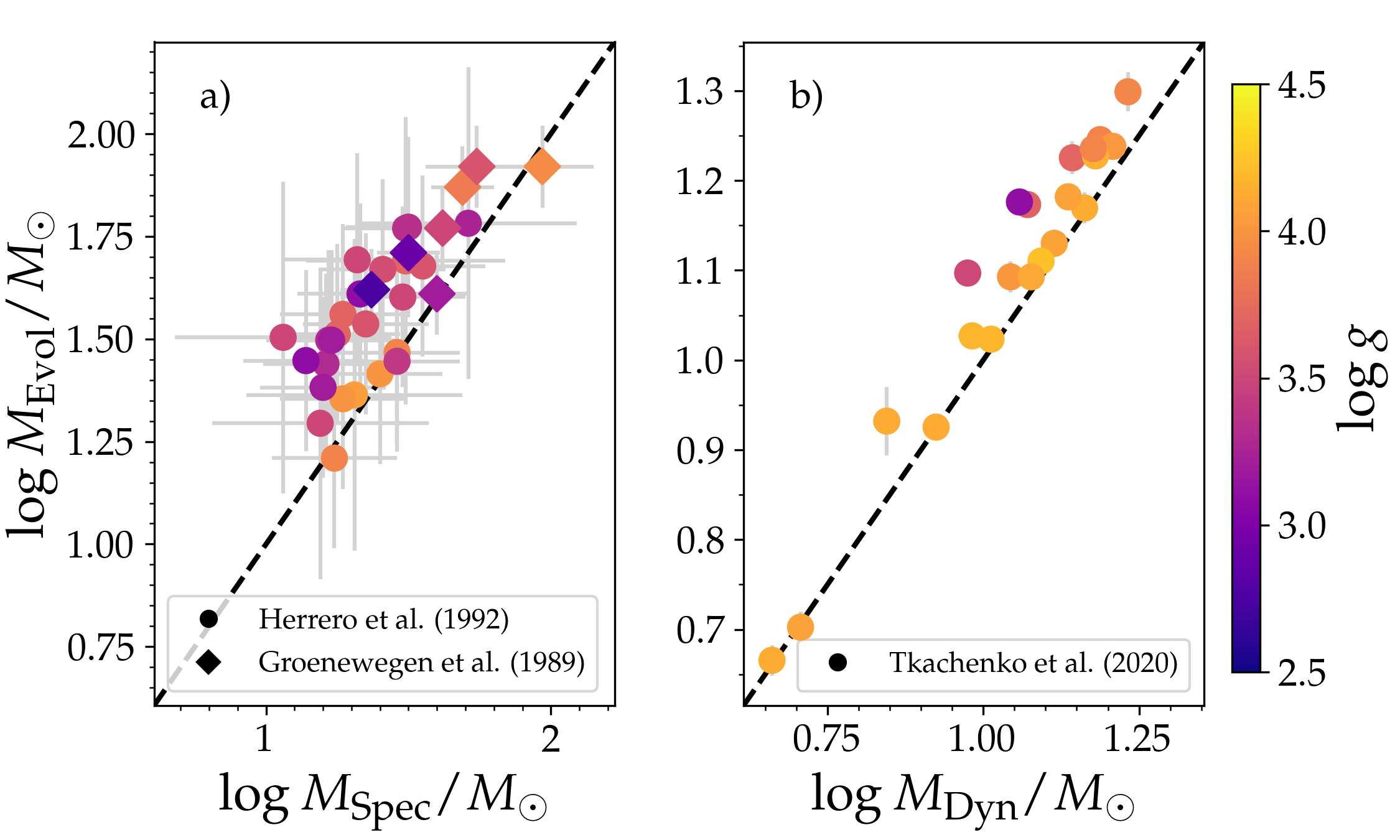}
\caption{Examples of observed mass discrepancies. Better agreement between both $M_\text{Evol}$-$M_\text{Spec}$ and $M_\text{Evol}$-$M_\text{Dyn}$ is generally found for less evolved stars. \textbf{a)} Observed mass discrepancy between spectroscopic (x-axis) and evolutionary (y-axis) masses derived for a sample of 32 high-mass stars \cite{Groenewegen1989A&A...221...78G,Herrero1992LNP...401...21H}. Similar errors to those for $M_\text{Spec}$ were assumed for $M_\text{Evol}$ for the round data points. \textbf{b)}~Observed mass discrepancy between dynamical (x-axis) and evolutionary (y-axis) masses derived for a sample of 11 double-lined eclipsing binaries \cite{Tkachenko2020A&A...637A..60T}. Errors on the measurements are typically smaller than the size of the data points. Figure made by the authors using data from \citet{Groenewegen1989A&A...221...78G}, \citet{Herrero1992LNP...401...21H}, and \citet{Tkachenko2020A&A...637A..60T}.
\label{fig:Mspec_vs_Mevol}}
\end{figure}

The mass discrepancy problem is a disagreement between spectroscopically derived stellar masses\footnote{The spectroscopic masses $M_\text{Spec}$ are obtained from the spectroscopic $\log g$ values in combination with radius estimates from, e.g., relations between $T_\text{eff}$ values, bolometric corrections, and spectral types \cite[see][]{Groenewegen1989A&A...221...78G}.} and those obtained from stellar evolution models. This disagreement appears on the HR diagram because stars and their expected evolutionary tracks do not overlap \cite{Groenewegen1989A&A...221...78G,Herrero1992LNP...401...21H}. Derived spectroscopic masses are systematically lower than evolutionary masses (see Figure~\ref{fig:Mspec_vs_Mevol} a)), hinting towards missing or inadequate physics in the standard stellar structure and evolution models used. In stellar binaries, the mass discrepancy is a disagreement between dynamically derived component masses and evolutionary masses from standard models when a common age is enforced (see Figure~\ref{fig:Mspec_vs_Mevol} b)). This problem was first seen for the binary systems SZ~Cen \cite{Gronbech1977A&A....55..401G,Andersen1990ApJ...363L..33A}, BW~Aqr \cite{Clausen1991A&A...246..397C}, and BK~Peg \cite{Clausen1991A&A...246..397C}, where the inclusion of CBM is needed in order to obtain satisfactory fits to the more massive components of the systems. 
These early studies did not consider a difference in CBM parameters between the components of the systems.

Large CBM parameters (primary $\alpha_\text{ov} = 0.3-0.5$; secondary $\alpha_\text{ov} = 0.1-0.4$) derived from vectors in the mass-luminosity plane were found for the high-mass ($M_1 = 39.5$\,M$_\odot$, $M_2 = 33.5$\,M$_\odot$) detached eclipsing binary system HD~166734 \cite{higgins_vink_2019}.
These stars are blue supergiants (BSGs), and explaining the population of BSGs is a long standing problem for stellar structure and evolution theory.
The area in the HR diagram where the BSGs are found is expected to be scarcely populated due to rapid post-MS stellar evolution, but the opposite is observed \cite{Fitzpatrick1990ApJ...363..119F}. 
There are two likely explanations for this: 1) the main-sequence is extended compared to standard models and BSGs are core-hydrogen-burning stars, or 2) BSGs are post-MS stars undergoing core helium burning. 
The mass-luminosity plane inference of high CBM parameters suggests that most blue supergiants (BSGs) are on the main-sequence close to the TAMS \cite{higgins_vink_2023}. 
However, the fact that BSGs are slow rotators ($v\sin i \leq 50$\,km\,s$^{-1}$) compared to hotter massive stars with $v\sin i \lesssim 400$\,km\,s$^{-1}$ seemingly supports the core helium burning scenario, because rotational velocity should decrease after the main-sequence as the stellar envelope expands \cite{Hunter2008A&A...479..541H}. 
This expansion is tied to the star's $\log g$ value, and \citet{Brott2011A&A...530A.115B} used $\log g$ values to calibrate the CBM parameter at 16\,M$_\odot$, finding $\alpha_\text{ov} = 0.335$. 
However, the drop from $v\sin i \approx 400$\,km\,s$^{-1}$ to $\leq 50$\,km\,s$^{-1}$ coincides with the effective temperature of $\approx 22000$\,K, where rotational braking due to enhanced mass-loss may occur \cite{Vink2010A&A...512L...7V}. 
Such braking requires a CBM parameter $\alpha_\text{ov} \geq 0.335$ to occur at masses as low as 10\,M$_\odot$.

Another binary system which suggests the importance of CBM is V380~Cyg. The primary component's mass discrepancy is extreme and may be in excess of $10-30\,\%$ \cite{Guinan2000ApJ...544..409G,Pavlovski2009MNRAS.400..791P,Tkachenko2014MNRAS.438.3093T,Tkachenko2020A&A...637A..60T}. 
One solution to this problem is to use a CBM parameter of $\alpha_\text{pen} = 0.6 \pm 0.1$ for the primary component, with no CBM required to reconcile the less evolved near-ZAMS secondary component \cite{Guinan2000ApJ...544..409G}. Recent updated component parameters show that a discrepancy also exists for the secondary component, which can be fixed by decreasing the metallicity and increasing the mass of the star within its $3\sigma$ error \cite{Tkachenko2014MNRAS.438.3093T}. For the primary a mass at the $3\sigma$ limit combined with a high rotation ($v_\text{ZAMS} = 241$\,km\,s$^{-1}$) and strong level of CBM ($\alpha_\text{pen} = 0.6$) was required to reconcile the evolutionary models with the observations \cite{Tkachenko2014MNRAS.438.3093T}. 
Another complication is that the primary has high microturbulence velocity ($\xi$$ = 15$\,km\,s$^{-1}$), and neglecting this in the spectroscopic analysis causes the effective temperature to be overestimated by $\approx 1700$\,K ($\approx 8\%$)\footnote{$\log g$ can be derived from the component masses and radii and was therefore held fixed for this comparison.} \cite{Tkachenko2020A&A...637A..60T}. 
This would likewise impact the derived $T_\text{eff}$ of the secondary both from spectroscopic disentangling and photometric analysis of the light curve. Appropriately accounting for the effect of microturbulence in the spectroscopic analysis in combination with the inclusion of CBM could fully explain the mass-discrepancy of this system. 
Such an analysis has yet to be carried out.

In comparison to the four binary systems discussed above, excellent fits to the observations were found for the binary systems V792~Her \cite{Fekel1991AJ....101.1489F}, AI~Phe \cite{Andersen1988A&A...196..128A}, and UX~Men \cite{Andersen1989A&A...211..346A} using standard models without CBM. 
These systems cover the mass range $1.2-1.5$\,M$_\odot$, whereas the more massive components of SZ~Cen, BW~Aqr, BK~Peg, and V380~Cyg have masses between $1.43-11.43$\,M$_\odot$. 
These seven systems provide some indication that a mass dependence may exist for CBM. 

\subsubsection{A search for mass dependent CBM using binary systems}\label{Sec:mass_dep_CBM}

Detached double-lined eclipsing binaries (DDLEB) provide great test beds for stellar structure and evolution models. The component masses, radii, and effective temperatures of DDLEBs can be precisely and accurately measured, and the components can reasonably be assumed to share a common age and composition. 
The advantage of using DDLEBs over single stars can clearly be seen from a comparison of the errors between panels a and b in Figure~\ref{fig:Mspec_vs_Mevol}. 

\begin{figure}[t]
\includegraphics[width=\linewidth]{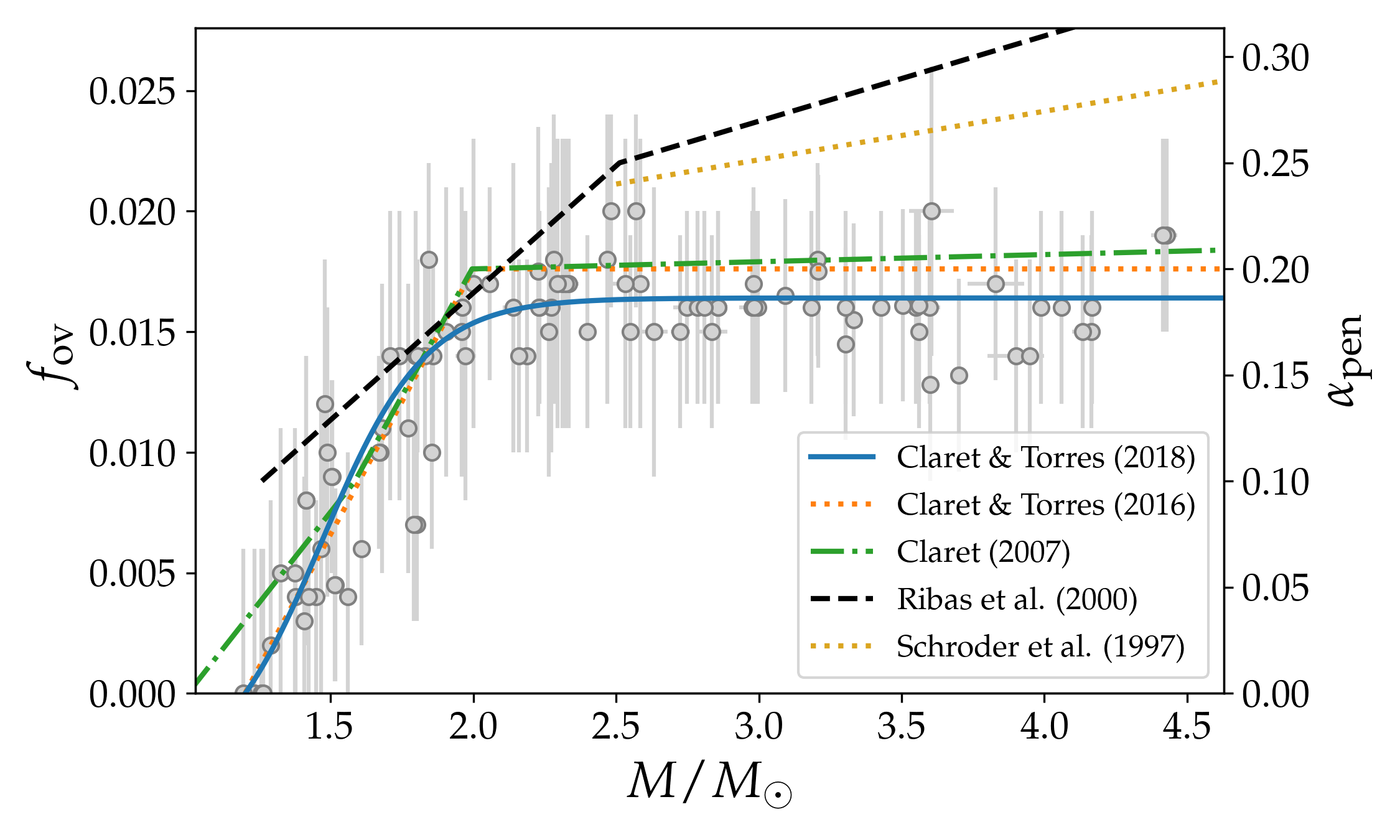}
\caption{Inferred mass versus CBM obtained for a sample of 50 DDLEBs (grey points with errors) \cite{Claret2019ApJ...876..134C}, compared to different empirical relations available in the literature obtained from modeling different samples of DDLEBs. Conversion factors of $\alpha_\text{pen} /f_\text{ov} = 11.36$ \cite{Claret2017ApJ...849...18C} and $\alpha_\text{pen} /\alpha_\text{ov} = 1$ were assumed for the inclusion of the individual curves in the figure. The plotted range is limited to the mass range of the 50 DDLEBs. Errors of $\sigma_{f_\text{ov}} = 0.006$ were assumed for main-sequence stars, while $\sigma_{f_\text{ov}} = 0.004$ were used for giants \cite{Claret2019ApJ...876..134C}. Figure made by the authors using data from \citet{Claret2019ApJ...876..134C}, 
\citet{Claret2018ApJ...859..100C}, 
\citet{Claret2017ApJ...849...18C}, \citet{Claret2016A&A...592A..15C}, \citet{Claret2007A&A...475.1019C}, \citet{Ribas2000MNRAS.318L..55R}, and \citet{Schroder1997MNRAS.285..696S}.
\label{fig:DLEB_fov_vs_mass}}
\end{figure}

The largest sample of DDLEBs that have been used to investigate the presence of a mass dependence of CBM consists of 50 systems (100 stars) in the mass range of 1.2 to 4.4\,M$_\odot$ \cite{Claret2016A&A...592A..15C}. For all of these systems, the masses and radii are known to a $3\%$ accuracy or better, while the effective temperatures are known to a $5\%$ accuracy. An ensemble study of these stars revealed that the extent of the CBM region appears to be steadily increasing with mass from $1.2-2$\,M$_\odot$ and reaches a plateau that persists up the upper limit of the mass in the sample of $4.4$\,M$_\odot$, see grey data points in Figure~\ref{fig:DLEB_fov_vs_mass}. The associated by-eye fit to the data \cite{Claret2018ApJ...859..100C} is shown by the solid blue line in Figure~\ref{fig:DLEB_fov_vs_mass}, while an earlier result by the same authors for a smaller sub-sample of 33 DDLEBs and where convective penetration was used instead of exponential diffusive overshoot is indicated by the orange dotted line \cite{Claret2016A&A...592A..15C}. 
The switch from using convective penetration to exponential diffusive overshoot was mainly a result of a change in the adopted stellar structure and evolution codes, and also allowed for derivation of a relation  $\alpha_\text{pen}/f_\text{ov} = 11.36 \pm 0.22$ \cite{Claret2017ApJ...849...18C} that could be used to perform a conversion between the two CBM parameters, see also Section~\ref{subsec:comp_params}. 
The small offset between the orange dotted and solid blue line in Figure~\ref{fig:DLEB_fov_vs_mass} for $M > 2$\,M$_\odot$ is caused by differences in the assumed primordial helium abundance \cite{Claret2017ApJ...849...18C}. 

One of the first studies of DDLEBs where a mass dependent CBM was investigated relied on a sample of three $\zeta$~Aurigae systems (wide eclipsing binaries where the primary component is a late-type bright giant or supergiant) and three related non-eclipsing binaries also containing an evolved primary component \cite{Schroder1997MNRAS.285..696S}. 
As indicated in Figure~\ref{fig:CBM_effects}, the effects of CBM on evolutionary tracks become more pronounced as stars age, so these systems were suggested as ideal test beds for CBM. The size of the CBM region was found to slightly increase with mass from $\approx 0.24\,H_\text{p,cc}$ at 2.5\,M$_\odot$ to $\approx 0.32\,H_\text{p,cc}$ at 6.5\,M$_\odot$, see yellow dotted curve in Figure~\ref{fig:DLEB_fov_vs_mass}. \citet{Ribas2000MNRAS.318L..55R} relied on a sample of eight DDLEBs with masses between 2 and 12\,M$_\odot$, and likewise found an increase in the extent of the CBM region with mass but with a steeper slope for the increase towards higher masses, see black dashed line in Figure~\ref{fig:DLEB_fov_vs_mass}. This latter result was largely guided by V380~Cyg, which was one out of only two systems in the sample with masses above 3.4\,M$_\odot$. Like the studies mentioned above, a large CBM parameter $\alpha_\text{ov} \approx 0.6$ was found for this system, whereas a lower $\alpha_{\rm ov} \sim$ 0.2-0.5 was needed for the similar mass system HV~2274. An age dependence of the CBM parameter was suggested as a possible solution to this difference in $\alpha_\text{ov}$. 
\citet{Ribas2000MNRAS.318L..55R} also used data from prior studies of lower mass stars for the construction of their mass versus CBM relation partly shown by the black dashed line in Figure~\ref{fig:DLEB_fov_vs_mass}. 
A similar study with a significant overlap (eight out of 13) in the considered sample of DDLEBs with masses between 1.35 and 27.27\,M$_\odot$ also arrived at a mass dependent CBM, but with a much shallower slope for stars with masses above $\approx 2$\,M$_\odot$ \cite{Claret2007A&A...475.1019C}, see green dashed-dotted curve in Figure~\ref{fig:DLEB_fov_vs_mass}. In this case the errors on the derived CBM parameters are large, and the CBM-Mass relation therefore ambiguous.

If BSGs are core-hydrogen-burning stars, their distribution in the HR diagram provides some evidence for a mass-dependent CBM. 
\citet{castro_etal_2014} provided the first observational spectroscopic HR diagram\footnote{In the spectroscopic HR diagram the luminosity is calculated as $\mathcal{L} = \frac{T_\text{eff}^4}{g}$, thereby becoming independent of distance and extinction measurements.} of massive stars in the Milky Way, and compared the main-sequence density distributions to non-rotating model grids with two values of the CBM parameter. They find evidence of the CBM increasing from $\alpha_\text{ov} = 0.1$ at 8\,M$_\odot$ to $\alpha_\text{ov} = 0.335$ at $\approx$15\,M$_\odot$, and larger $\alpha_{\rm ov}$ is needed for higher masses. The inclusion of rotation in their models still showed that CBM is required, but mass-dependence is not unambiguous.

\subsubsection{Evidence against mass dependent CBM and complications}

The detection of a mass dependent CBM relying on ensembles of DDLEBs is not unambiguous and the reliability of the result has been questioned in several cases. 
\citet{Costa2019MNRAS.485.4641C} studied an earlier sample of 38 out of the 50 DDLEBs mentioned above, using stellar models including both CBM and rotational mixing and applied a Bayesian analysis to investigate the mass dependence of CBM \cite{Costa2019MNRAS.485.4641C}. 
Due to the wide scatter and large errors on the derived CBM parameters, the authors do not find a clear mass dependence on the CBM but rather identify a wide distribution of valid CBM parameters between $\approx 0.15-0.4\,\text{H}_{p,cc}$ for $M > 1.9$\,M$_\odot$, contrary to the constant value shown for the solid blue and dotted orange curves in Figure~\ref{fig:DLEB_fov_vs_mass}. 
They suggest that the distribution could be explained using models with a constant CBM parameter $\alpha_{\rm ov} = 0.2$ and initial rotational velocities between 0-80\% break-up velocity. 
\citet{Constantino2018A&A...618A.177C} analyzed a different subset of eight representative systems out of the 50 DDLEBs to determine whether or not the derived CBM parameters for each system are unique. 
They found that the uncertainties on the derived $f_\text{ov}$ values are high and a single value of the CBM parameter could be used for the entire mass range of $1.3-3.7$\,M$_\odot$. No mass dependence on the CBM was found, but the results did indicate that CBM was needed for $M > 2$\,M$_\odot$. The derived uncertainties may be too pessimistic, as this work did not take into account additional constraints available from including the effective temperatures in the analysis \cite{Claret2019ApJ...876..134C}. 
Uncertainties could be further reduced if more precise effective temperatures and metallicities were obtained \cite{Constantino2018A&A...618A.177C}. 

\citet{Meng2014ApJ...787..127M} considered a sample of four eclipsing binary systems relying on a CBM formalism that did not use the pressure scale height to determine the extent of the CBM region, and found no CBM dependence on the mass in the range $1.2$ to $2.5$\,M$_\odot$. 
\citet{Stancliffe2015A&A...575A.117S} also found no CBM dependence on mass or metallicity for their sample of nine eclipsing binaries with $M = 1.3- 6.2$\,M$_\odot$ using both exponential diffusive overshoot and a nontraditional CBM formalism where an adjustment was made directly to the Schwarzschield criterion based on the ratio between the radiation and gas pressure and a free CBM parameter \cite{Stancliffe2015A&A...575A.117S}. 
A more recent study relied on a sample of 11 DDLEBs with $M = 4.6-17.1$\,M$_\odot$ where the fundamental and atmospheric parameters were all derived using the same methodology, in contrast to the ensemble studies mentioned above \cite{Tkachenko2020A&A...637A..60T}. No mass dependence on the CBM parameter was found for this sample, but the observed mass discrepancy could be explained by a combination of a need for higher core masses and the lack of proper treatment of microturbulent velocities in the spectroscopic analysis of the stars.  

Discussions dating back more than 30 years argue that, in order to constrain helium abundances, opacity tables, mixing length and/or CBM parameters, the binary star parameters must be known to accuracies of at least 1\% (for radii), 2\% (for mass and temperature), and 25\% (for metallicity) \cite{Andersen1991A&ARv...3...91A}. 
These numbers have been backed up by more recent statistical studies on using binary systems to constrain CBM. 
\citet{Valle2016A&A...587A..16V} argue that systems with masses between 1.1 and 1.6\,M$_\odot$ where both components are on the main-sequence cannot be used to calibrate CBM when the errors on the component masses are appropriately accounted for. 
The only exception is when the primary is in the last 5\% of its main-sequence evolution. 
This study assumed errors on the effective temperature, metallicity, component masses, and radii of 100\,K, 0.1\,dex, 1\%, and 0.5\% respectively. 
The observed biases and uncertainties are reduced when the errors on observed parameters are reduced. In comparison, a similar statistical analysis using the more evolved binary system TZ~Fornacis ($M_\text{primary} = 2.057\pm0.001$\,M$_\odot$, $M_\text{secondary} = 1.958\pm0.001$\,M$_\odot$) found that good constraints on CBM could be obtained, owing to the low errors on the component masses \cite{Valle2017A&A...600A..41V}. 
Biases were also found from a subsequent statistical analysis assuming a $M=2.50$\,M$_\odot$ primary star in three different post-main-sequence evolutionary stages and a $M=2.38$\,M$_\odot$ secondary \cite{Valle2018A&A...615A..62V}. 
However, differentiating between cases of no and mild CBM is generally possible, unless the primary is undergoing central helium burning. For the 50 DDLEBs in Figure~\ref{fig:DLEB_fov_vs_mass}, 59\% of the component masses are known to better than a 1\% accuracy, while for 20\% and 2\% of the stars the masses are known to a 0.5\% and 0.1\% accuracy, respectively. 
The errors on the component masses were not taken into account when deriving the CBM parameters for the 50 DDLEBs in Figure~\ref{fig:DLEB_fov_vs_mass}, but the metallicity was allowed to vary within the observed errors \cite{Claret2017ApJ...849...18C,Claret2018ApJ...859..100C,Claret2019ApJ...876..134C}.

The results of the statistical analyses mentioned above seem backed up by earlier attempts at studying CBM using 49 DDLEBs, which found that both models with and without CBM provide satisfactory fits to the observations of 80\% of the systems, but models with CBM provide better fits for systems with components in the post-main-sequence stage of stellar evolution \cite{Pols1997MNRAS.289..869P}. This once again points towards the complications arising from using main-sequence DDLEBs to constrain CBM. 

Finally, all of the studies discussed above have been done under the same assumption that the evolution of the binary systems can be treated as the evolution of two single stars without accounting for impacts from binary interactions on the stellar models. Such an assumption has been shown to be reasonable for the evolution of well-detached preinteraction binary systems at least when rotational mixing in single versus binary stars is considered \cite[e.g.][]{Mahy2020A&A...634A.118M}. 
For detached, short-period binaries with strong tidal interactions, some studies suggest an enhanced rotation mixing in the presence of strong dynamical tides \cite{deMink2009A&A...497..243D}. Others suggest that the effects of tides are limited in detached systems \cite{Mahy2020A&A...634A.118M,Martins2017A&A...607A..82M}, while another recent study indicate that internal mixing is less efficient in detached binary systems than for single stars \citep{Pavlovski2023arXiv230104215P}.


\subsection{Asteroseismology}
\label{subsec:asteroseismology}

Asteroseismology is the study and interpretation of stellar pulsations, and provides a powerful tool for studying stellar interiors. 
The pulsations are observed as variations in the surface brightness of the stars and extend deep into the stellar interiors, thereby carrying information about the conditions within. 
Modifications to the interior structure result in changes to the expected oscillation frequencies, and  confronting predicted oscillation frequencies with observations in a process known as asteroseismic modeling provides important constraints on stellar structure and evolution theory. 
Such constraints have especially been made possible in the past couple of decades since the advent of space telescopes including WIRE (e.g.,~\cite{Buzasi2000ApJ...532L.133B}), MOST \cite{Walker2003PASP..115.1023W}, CoRoT \cite{Auvergne2009,Baglin2009}, Kepler \cite{Borucki2010,Koch2010}, K2 \cite{Howell2014PASP..126..398H}, BRITE \cite{Weiss2014}, and TESS \cite{Ricker2015}, which provide high-precision, high-cadence, and long time base-line photometric light curves. 
This high-quality data resulted in a drastic increase in the number of detected oscillation frequencies in stars all across the HR diagram and provided the high frequency resolution needed for mode identification required for asteroseismic modeling. 
Several review papers and books on asteroseismology already exist (e.g.,~\cite{Aerts2010aste.book.....A,Chaplin2013ARA&A..51..353C,Aerts2019ARA&A..57...35A,Bowman2020FrASS...7...70B,Holdsworth2021FrASS...8...31H,Christensen-Dalsgaard2021LRSP...18....2C,Aerts2021RvMP...93a5001A,Guzik2021FrASS...8...55G,Kurtz2022ARA&A..60...31K}, and references therein). Therefore, we refer to these papers for details on different types of pulsators, analysis methods, new discoveries, and procedures for asteroseismic modeling, and focus here on the inferences made on the CBM from asteroseismic studies of both single and binary stars.

\subsubsection{Onset of the convective core}

One avenue to place constraints on CBM is studying stars in the mass regime where the transition between radiative to convective cores is expected to occur ($\approx 1.1$\,M$_\odot$). 
In this regime we find the solar-like oscillators that oscillate in pressure (p) modes. 
These oscillations are an excellent probe of the convective core size, because they are modified by the presence of acoustic glitches in the sound speed profile caused by the sharp chemical gradient at the convective boundary \cite{Houdek2007MNRAS.375..861H,Roxburgh2007MNRAS.379..801R,Cunha2007ApJ...666..413C,Cunha2011A&A...529A..10C}. As an example, a convective core with either little or a moderate amount of overshooting was found for the $1.18\pm0.04$\,M$_\odot$ star KIC~12009504 whereas no convective core could unambiguously be found for KIC~6106415 ($1.11\pm0.04$\,M$_\odot$) \cite{SilvaAguirre2013ApJ...769..141S}. 
Another example is HD~203608 ($0.94\pm 0.09$\,M$_\odot$), for which models with convective cores agree better with the observations than models without \cite{Deheuvels2010A&A...514A..31D}.

Stars like the Sun arrive at the zero-age main-sequence (ZAMS) with a small convective core, which quickly disappears during the main-sequence evolution. 
The presence of the convective core is caused by an excess of $^3$He and $^{12}$C at the ZAMS, which are transformed to $^4$He and $^{14}$N through highly exothermic nuclear burning capable of sustaining a convective core \cite{Roxburgh1985SoPh..100...21R}. 
The inclusion of CBM extends the lifetime of the convective cores \cite{Roxburgh1985SoPh..100...21R}, possibly even until the end of the main-sequence if sufficiently high CBM parameters are considered \cite{Mowlavi1993ASPC...40..454M} (see also \citet{Roxburgh1985SoPh..100...21R} for a detailed discussion). For HD~203608, such overshooting ($\alpha_\text{ov} = 0.17\pm 0.03$) allows the convective core to survive until the present age of the star \cite{Deheuvels2010A&A...514A..31D}. For this particular star, the convective core would have disappeared at an age of 200\,Myr without CBM, whereas it is expected to survive until $\approx 7$\,Gyr with overshooting. Stars of similar masses where no convective cores are found are of equal interest as they provide an upper limit for the extent of the CBM region. 

\subsubsection{Extent of the CBM region across the main-sequence}

\begin{figure}
\includegraphics[width=\linewidth]{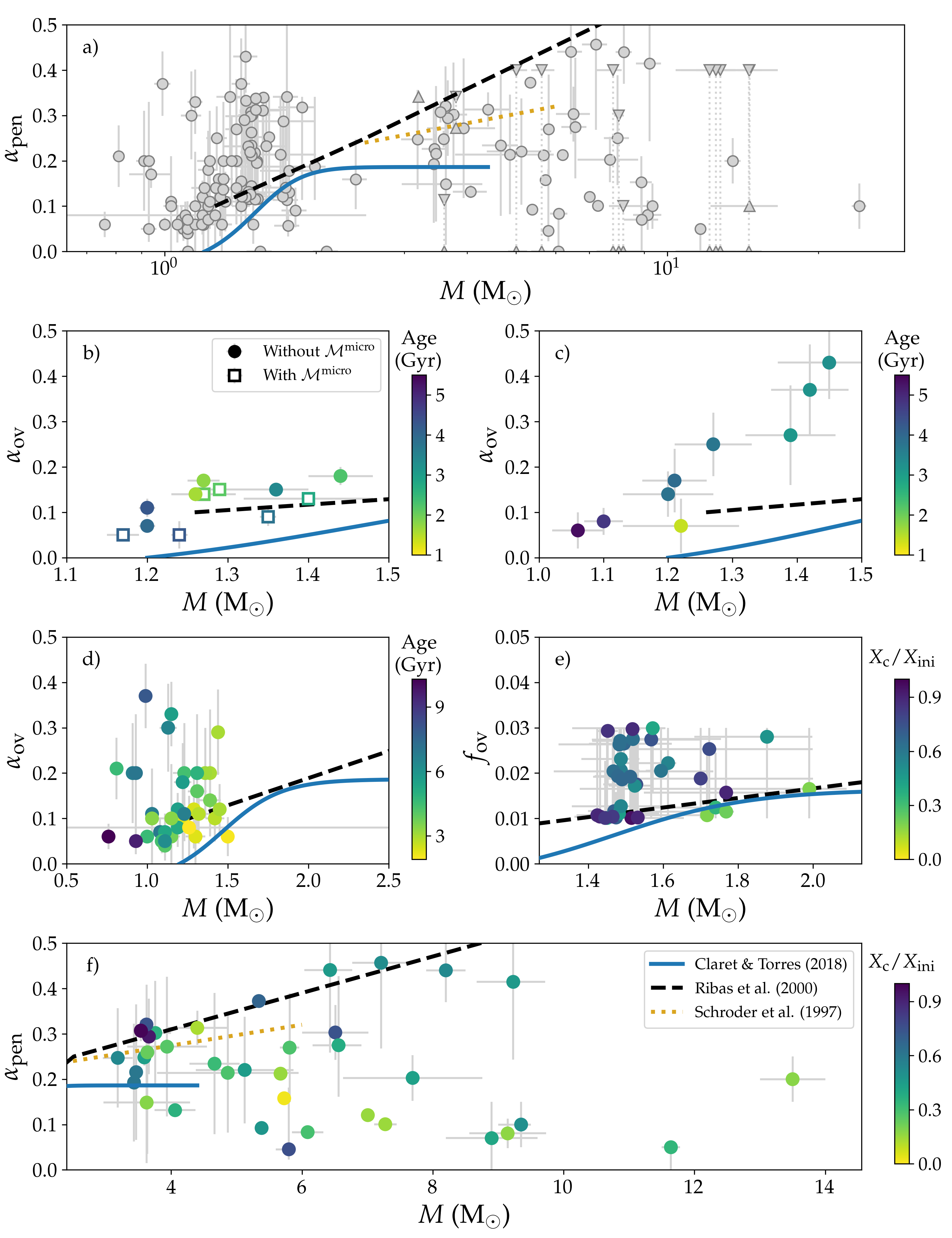}
\caption{
    Asteroseismically derived CBM parameters available in the literature for 154 pulsating main-sequence stars.
    Panel a displays the results from all studies we identified in the literature \cite{Claret2018ApJ...859..100C,Ribas2000MNRAS.318L..55R,Schroder1997MNRAS.285..696S,Dupret2004A&A...415..251D,Ausseloos2004MNRAS.355..352A,Mazumdar2006A&A...459..589M,Briquet2007MNRAS.381.1482B,Desmet2009MNRAS.396.1460D,Deheuvels2010A&A...514A..31D,Aerts2011A&A...534A..98A,Briquet2011A&A...527A.112B,Briquet2012MNRAS.427..483B,SilvaAguirre2013ApJ...769..141S,Walczak2013MNRAS.432..822W,Moravveji2015A&A...580A..27M,Moravveji2016ApJ...823..130M,Deheuvels2016A&A...589A..93D,Bellinger2016ApJ...830...31B,Schmid2016A&A...592A.116S,Daszynska-Daszkiewicz2017MNRAS.466.2284D,SanchezArias2017A&A...597A..29S,Buysschaert2018A&A...616A.148B,Aerts2019A&A...624A..75A,Khalack2019MNRAS.490.2102K,Hendriks2019PASP..131j8001H,Wu2019ApJ...881...86W,Mozdzierski2019A&A...632A..95M,Fedurco2020A&A...633A.122F,Wu2020ApJ...899...38W,Viani2020ApJ...904...22V,Mombarg2021A&A...650A..58M,Michielsen2021A&A...650A.175M,Pedersen2021NatAs...5..715P,Pedersen2022ApJ...930...94P,Niemczura2022MNRAS.514.5640N,Szewczuk2022MNRAS.511.1529S}.
Panels b-d show sub-samples of the stars from panel a, where an ensemble study was carried out using the same stellar structure and evolution code and modeling methodology \cite{Deheuvels2016A&A...589A..93D,Viani2020ApJ...904...22V,Bellinger2016ApJ...830...31B,Mombarg2021A&A...650A..58M}.
In panel f only $M \geq 3$\,M$_\odot$ stars from panel a with a known $X_\text{c}/X_\text{ini}$ value are shown. For stars without estimated errors on the initial mass and CBM parameter, we adopt the average fractional errors from the sample with error estimates. 
 To combine the data from all sources in panel a and f we assumed $\alpha_\text{ov} \approx \alpha_\text{pen}$ and $11.36 f_\text{ov} \approx \alpha_\text{pen}$. 
 Figure made by the authors.
\label{fig:asteroseis_cbm_vs_mass}}
\end{figure}

To compare asteroseismic inferences on CBM to the results presented in Figure~\ref{fig:DLEB_fov_vs_mass} for the study of binary stars, we compiled a sample of pulsating main-sequence stars for which a mass and CBM parameter has been derived asteroseismically. This full sample is shown in panel a of Figure~\ref{fig:asteroseis_cbm_vs_mass}, including also three of the CBM--mass relations from Figure~\ref{fig:DLEB_fov_vs_mass} for reference. The asteroseismic CBM parameters were all converted to $\alpha_\text{pen}$ assuming $\alpha_\text{pen} \approx \alpha_\text{ov}$ and $\alpha_\text{pen} \approx 11.36\, f_\text{ov}$. The lack of measurements between $\approx 2$ and $\approx 3$\,M$_\odot$ corresponds to the gap between the $\delta$~Sct and SPB instability strips. For 13 stars in this sample only a lower (upright triangle) and upper (inverted triangle) limit on the CBM are available. Nine of the 13 stars are located in the open cluster NGC~6910 \cite{Mozdzierski2019A&A...632A..95M}. 
As seen in panel a, there is a general trend of increasing CBM parameter with increasing stellar mass for stars with $M \lesssim 2.5$\,M$_\odot$\footnote{The Spearman's rank correlation coefficient for this sub-sample is 0.324 with a $p$-value of 0.0011, corresponding to strong evidence for a positive correlation between the CBM parameter and stellar mass.}, however, the scatter is larger and the CBM parameters generally higher than the values found for the sample of 50 DDLEB in Figure~\ref{fig:DLEB_fov_vs_mass}. For the higher mass stars, a wide range in CBM paramaters are found with no clear dependence on stellar mass and inconsistent with a single value of the CBM parameter as previously found for the DDLEBs. We note that for a few of the stars, no CBM was needed to match the models to the observations.

The CBM parameters shown in Figure~\ref{fig:asteroseis_cbm_vs_mass} were derived using a variety of different stellar structure and evolution codes and different 1D prescriptions for the CBM. 
As discussed in Section~\ref{subsec:cbm_limits} and \ref{subsec:comp_params}, making direct comparisons and conversions between CBM parameters obtained from different sources is non-trivial. 
For this reason, we show in panels b-e four different subsamples of the stars in panel a where a consistent stellar structure and evolution code and modeling methodology was applied to each sample. 
The data are color-coded according to the age of the stars either measured in Gyrs or taken as the ratio between the current and initial core hydrogen mass fraction $X_\text{c}/X_\text{ini}$. 
In panel f, we show all stars with $M \geq 3$\,M$_\odot$ for which an estimate of $X_\text{c}/X_\text{ini}$ is available irrespective of the adopted code and modeling procedure. As the main-sequence lifetime is largely dependent on both the initial stellar mass and internal mixing properties of the stars, $X_\text{c}/X_\text{ini}$ provides a much better age indicator when doing comparison studies \cite{Pedersen2022ApJ...940...49P}. 
For this high-mass sample, we expect the differences in CBM prescription to be largely insignificant (cf. Section~\ref{subsec:cbm_limits}). 

Panels b and c of Figure~\ref{fig:asteroseis_cbm_vs_mass} show two different ensemble studies of solar-like oscillating stars, which both indicate an increase in the CBM parameter with mass albeit at different levels. For both samples, the derived CBM parameters are higher than the relation derived by Claret \& Torres \cite{Claret2018ApJ...859..100C} from DDLEBs. The sample in panel b was modeled using the \texttt{Cesam2k} code both with (open squares) and without (filled circles) microscopic diffusion, indicating that less CBM is needed to explain the observations if microscopic diffusion is taken into account \cite{Deheuvels2016A&A...589A..93D}. 
This is in line with previous results from the open cluster M67 \cite{Michaud2004ApJ...606..452M,Viani2017EPJWC.16005005V}. 
In comparison, the sample in panel c was modeled using the \texttt{YREC} code, using the concept of an effective overshoot parameter which accounts for the automatic adjustment of the size of the CBM region to avoid nonsensical core sizes \cite{Viani2020ApJ...904...22V}. A similar study also derived the effective overshoot parameters for a sample of solar-like oscillators, finding the current effective overshoot parameter to be zero up to $\approx 1.1$\,M$_\odot$ above which a scatter in the effective $\alpha_\text{ov}$ values appears and increases towards higher masses \cite{Angelou2020MNRAS.493.4987A}. 

Both of the samples of pulsating stars in panels d and e were modeled using different versions of the \texttt{MESA} code and using machine learning algorithms to derive the stellar parameters from the observed oscillation properties of the stars. For the sample of solar-like stars in panel d, the majority of the stars show an increase in CBM parameter with mass and generally higher CBM parameters than the relation predicted by Claret \& Torres \cite{Claret2018ApJ...859..100C}. For the sample of $\gamma$~Dor stars in panel e), no clear mass dependence is found. However, the errors on the CBM parameter are large and span the entire parameter range considered \cite{Mombarg2021A&A...650A..58M}. This is likely caused by the inability of the neural network to capture the fine details of the pulsation properties of the models which are important for constraining this parameter.

The final panel f of Figure~\ref{fig:asteroseis_cbm_vs_mass} focuses on the $M \geq 3$\,M$_\odot$ stars in the sample where a common $X_\text{c}/X_\text{ini}$ age indicator is available. For 26 of the stars shown here an ensemble study was carried out, whereas the remaining stars were studied individually. As already mentioned earlier, a much larger scatter in the CBM parameter is seen here with no indication that a single parameter fits all of the stars. No clear dependence on age is seen either. One possible explanation for the observed scatter is that additional interior physical processes are influencing the measured and required CBM. As an example, envelope mixing arising from, e.g., internal rotation has a similar effect in bringing additional hydrogen to the core and can therefore enhance the effective core size. This gives rise to some degeneracies between the CBM and envelope mixing, which can be difficult to disentangle. On the other hand, internal magnetic fields can inhibit CBM. As an example, the low amount of overshooting ($f_\text{ov}= 0.004_{-0.002}^{+0.012}$) found for the pulsating magnetic B-type star HD~43317 suggests that the magnetic fields of the star might be suppressing the mixing near the core boundary \cite{Buysschaert2018A&A...616A.148B}. The lack of observed high-radial order gravity (g) modes suggests that a near-core magnetic field of at least 500\,kG is suppressing these modes in the star \cite{Lecoanet2022MNRAS.512L..16L}. Prior to these results, the asteroseismic modeling of the magnetic $\beta$~Cep star V2052 Ophiuchi revealed a lower overshoot parameter ($\alpha_\text{ov} = 0.07_{0.07}^{0.08}$,~\cite{Briquet2012MNRAS.427..483B}) compared to its similar mass $\beta$~Cep counterpart $\theta$~Ophiuchi ($\alpha_\text{ov} = 0.44 \pm 0.07$,~\cite{Briquet2007MNRAS.381.1482B}), indicating that magnetic fields are likely inhibiting the CBM for this star as well. 
For the lower mass ($M=2.4$\,M$_\odot$) magnetic pulsating Ap star, a CBM parameter of $f_\text{ov} = 0.014$ was found \cite{Khalack2019MNRAS.490.2102K} with no clear indication of inhibition of the CBM by the magnetic fields of the star. 

\subsubsection{Differentiating between different CBM prescriptions}

Gravity mode oscillators are of particular interest for studying CBM, as g~modes have their main probing power in the near core regions of the stars. 
SPB stars ($3-10$\,M$_\odot$, g-modes) have been shown to be capable of distinguishing not only between different shapes of $D_\text{CBM} (r)$ profiles \cite{Pedersen2018A&A...614A.128P,Michielsen2019A&A...628A..76M} but also between different choices of temperature gradients in the CBM region \cite{Michielsen2019A&A...628A..76M}. A higher frequency precision corresponding to longer light curves ($>1$\,yr) are needed to distinguish between different shapes of the mixing profiles compared to between radiative and adiabatic temperature gradients ($>90$\,d) \cite{Michielsen2019A&A...628A..76M}. 
Oscillations in $\beta$~Cep variables ($10-25$\,M$_\odot$, p- and g-modes) can also distinguish between different CBM profiles and temperature gradients, but require the mass of the star to be known to at least $1\,\%$ precision. Similar theoretical studies have not yet been carried out for $\gamma$~Dor (g-modes) or $\delta$~Sct (p-modes) stars, while it was found for solar-like oscillators that their p-modes 
could not be used to differentiate between exponential diffusive overshoot and step overshoot \cite{Deheuvels2016A&A...589A..93D}.

So far asteroseismic inferences on the shape of $D_\text{CBM} (r)$ and choice of $\nabla_T$ have only been attempted in a few cases, with the majority of studies focusing on estimating "just" the extent of the CBM region as discussed in the previous section. KIC~10526294 and KIC~7760680 are the first two SPB stars where a comparison between results assuming a step and exponential diffusive overshoot prescription was made \cite{Moravveji2015A&A...580A..27M,Moravveji2016ApJ...823..130M}. For both stars it was found that models with exponential diffusive overshoot provide better matches to the observed period spacing pattern compared to models with step overshoot. In a subsequent study of KIC~7760680, the CBM profile was fixed to that of an extended convective penetration while the temperature gradient was varied between a purely radiative gradient and one gradually changing from adiabatic to radiative based on the Peclet number (see Section~\ref{subsec:ext_conv_pen}) 
\cite{Michielsen2021A&A...650A.175M}. For this star, a radiative temperature gradient in the CBM region was preferred. However, it was also found that models without CBM were statistically preferred over models including CBM when the number of additional free parameters was taken into account. A comparison between step and exponential overshoot was briefly made for the SPB star KIC~8264293. The difference between the two was found to be minor, likely because the star is near the ZAMS and thereby lacks a significant chemical gradient at the core boundary, and the more extensive asteroseismic modeling of the star focused only on the exponential diffusive overshoot \cite{Szewczuk2022MNRAS.511.1529S}. 
Finally, an ensemble asteroseismic study of 26 SPB stars revealed that 54.9\% of the stars preferred convective penetration, whereas for 45.1\% the exponential diffusive overshoot prescription did better at reproducing the observed period spacing pattern \cite{Pedersen2021NatAs...5..715P,Pedersen2022ApJ...930...94P}. 


\section{Discussion \& Future Work}
\label{sec:discussion}

In this review, we provided an overview of convective boundary mixing (CBM) in the main sequence stars.
We discussed the most frequently-used prescriptions for including CBM in 1D stellar evolution models.
We described CBM from a hydrodynamics perspective, with an emphasis on lessons learned from simulations.
We provided an overview of the observations that are at odds with ``standard'' 1D models, and showed how excess (often mass-dependent) mixing can better align models and observations.

Despite great progress in recent years, there remains plenty of work to do before a complete understanding of CBM will be achieved.
In particular, most CBM processes described hydrodynamically and in simulations still lack robust parameterizations in 1D models.
We encourage 1D modelers and 3D numericists to forge partnerships to derive, test, and apply new simulation-based prescriptions.
We note in particular that there is a great deal of degeneracy in the \emph{language} which is used to describe CBM processes in the literature (e.g., ``overshoot'' and ``penetration'' are often used interchangeably), and our community must adopt language which clearly differentiates between the various physical mechanisms at work.
Below, we enumerate suggestions for future work and recommendations for future experiments which we believe will help to sort out the decades-old problem of CBM in stellar evolution.

From the perspective of 1D modeling and applying those models to observations, we have the following suggestions:
\begin{enumerate}
    \item First and foremost, it is valuable to ``grow the catalogue'' of CBM observations.
    More observational constraints will allow us to not only test and verify new models but also may allow us to understand how complications such as e.g., rotation affect CBM.
    \item A uniform analysis of past observations using a consistent stellar structure and boundary mixing scheme should be performed.
    \item To ease comparisons in future work, authors should clearly state which quantities their CBM prescriptions mix. 
    Specifically, does CBM adjust $\grad_\text{T}$ or not? 
    \item Evidence for extended convective penetration (Section~\ref{subsec:ext_conv_pen}) is seen in hydrodynamical simulations, and this prescription should be included in more stellar structure codes and models.
    \item In main sequence intermediate- to high-mass stars, the mass and radius of the convective core should be clearly reported along with the mass and size of the CBM region.
    Whether the reported convective core mass does or does not include the mass in the CBM region should also be specified. This circumvents difficulties associated with making comparisons between codes using different CBM prescriptions and methods of limiting the size of the CBM region.
    \item When reporting ages of stars on the main sequence, also report a quantity such as the core hydrogen fraction $X_c/X_{\rm init}$ for easier comparison across works.
\end{enumerate}

We also recommend that the following experiments be performed and prescriptions be derived from 3D hydrodynamical simulations:
\begin{enumerate}
    \item Whenever possible, 3D hydrodynamical simulations should strive to provide prescriptions that do not have free parameters but instead rely on stellar structure.
    \item For example, overshoot depth and turbulent diffusive mixing profiles, should be carefully calibrated and parameterized so that overshoot can be evaluated as a function of stellar structure rather than a specified $f_{\rm ov}$.
    \item 1D prescriptions derived from 3D simulation data should be validated using the same initial conditions employed in the 3D simulations. If the 1D prescription produces a different result from the 3D data, this should be explored in detail.
    \item Simulations probing the thermal structure near a convective boundary should be evolved until thermal equilibrium is achieved. Performing short simulations which are initialized with CBM regions of various sizes can however qualitatively answer the question, ``which way does the convective boundary move?''
    \item It is not clear how to properly parameterize dissipation, but dissipation sets the size of a convective penetration region. Future studies should answer the following: what sets the magnitude of the viscous dissipation? How does rotation affect it? How does magnetism and the presence of Ohmic dissipation affect it?
    \item Entrainment is important when convective regions are first forming, or when the convective luminosity or nuclear burning change rapidly compared to the convective overturn timescale. 
    These evolutionary stages should be modeled by time-dependent convection (TDC) prescriptions \citep{jermyn_etal_2022_mesa6}. 
    Future work should test whether TDC models reproduce the entrainment rates at convective boundaries observed in simulations, and TDC models should be improved where they disagree with simulations.
\end{enumerate}

A long-standing problem for stellar modelers and observers is the uncertainty that convective boundary mixing introduces into 1D stellar models.
The uncertainties associated with CBM not only affect studies focused on stars, but also ripple through other astrophysical disciplines which depend on state-of-the-art stellar models.
Great strides have been made in the past few decades in understanding CBM both from observations, 1D, and 3D simulations.
By combining the efforts of these often disparate lines of work, we can create and validate new mixing prescriptions and solve this long-lived problem.

\vspace{6pt} 



\authorcontributions{E.H.A and M.G.P. contributed equally in all aspects to this manuscript and have read \& agreed to the published version of the manuscript.}

\funding{
E.H.A. was supported by a CIERA Postdoctoral Fellowship.
This research was supported in part by the National Science Foundation under Grant No. NSF PHY-1748958 as well as through the TESS Guest Investigator program Cycle 4 under Grant No.\ 80NSSC22K0743 from NASA and  by the Professor Harry Messel Research Fellowship in Physics Endowment, at the University of Sydney (M.G.P.).}

\dataavailability{Inlists used to generate MESA stellar models in this work and data plotted in original figures are available online in the supplementary materials Zenodo \citep{supp}.}

\acknowledgments{
EHA thanks Adam Jermyn, Daniel Lecoanet, Benjamin Brown, Jeffrey Oishi, Adrian Fraser, and Rafa Fuentes for years of discussions on topics related to CBM.
We thank Raphael~Hirschi, Casey~Meakin, Federico~Rizzuti, and Tami~Rogers for allowing us to reproduce figures from their past work.
We also thank Jorick Vink, Dominic Bowman, and Jennifer van Saders for inviting and organizing us in writing this review.
We are grateful to the participants of KITP's ``Probes of Transport in Stars'' program from Fall 2021, where we had many useful discussions regarding CBM.
}

\conflictsofinterest{The authors declare no conflict of interest. The funders had no role in the design of the study; in the collection, analyses, or interpretation of data; in the writing of the manuscript; or in the decision to publish the~results.} 



\abbreviations{Abbreviations}{
The following abbreviations are used in this manuscript:\\

\noindent 
\begin{tabular}{@{}ll}
BSG & Blue supergiant\\
CBM & Convective Boundary Mixing\\
CZ & Convection Zone\\
DDLEB & Detached double-lined eclipsing binary\\
eMSTO & extended main-sequence turn-off\\
LHS & Left-hand side\\
RHS & Right-hand side\\
RZ & Radiative Zone\\
PZ & Penetrative Zone\\
TAMS & Terminal age main sequence\\
ZAMS & zero-age main sequence
\end{tabular}
}

\appendixtitles{no} 
\appendixstart
\appendix



\begin{adjustwidth}{-\extralength}{0cm}

\reftitle{References}


\bibliography{references.bib,penconv.bib,schwarzledoux.bib}

\end{adjustwidth}
\end{document}